\documentclass[aps,pre,floats,floatfix,twocolumn]{revtex4-1}

\usepackage{graphicx}
\usepackage{epstopdf}
\graphicspath{{./Figs/}}

\usepackage{latexsym}
\usepackage{amsmath}
\usepackage{amssymb}
\usepackage{bm}
\usepackage{wasysym}

\usepackage{ifthen}
\usepackage[usenames, dvipsnames]{color}
\usepackage[colorlinks,allcolors=blue,bookmarks=true]{hyperref}

\newcommand{\trc}{\mbox{trace}}

\newcommand{\amatrix}[1]{\begin{matrix} #1 \end{matrix}}

\newcommand{\ket}[1]{\left| #1 \right\rangle}
\newcommand{\braket}[1]{ \left\langle #1 \right\rangle}

\newcommand{\BraKet}[3]{ \left\langle #1 \middle| #2 \middle| #3 \right\rangle}

\newcommand{\be}[1]{\begin{eqnarray}{\label{e#1}}} 
\newcommand{\beq}{\begin{eqnarray}}
\newcommand{\eeq}{\end{eqnarray}} 

\newcommand{\hide}[1]{}
\newcommand{\Eq}[1]{\textcolor{blue}{{Eq.}\!~(\ref{#1})}} 
\newcommand{\Ap}[1]{\textcolor{blue}{{Appendix}\!~\ref{#1}}} 
 
\newcommand{\Fig}[1]{\textcolor{blue}{Fig.}\!~\ref{#1}}

\newcommand{\hrefl}[2]{\href{#2}{(#1)}}

\begin{document}

\title{Semiclassical theory of strong localization for quantum thermalization}

\author{Christine Khripkov$^{1}$, Amichay Vardi$^{1}$, Doron Cohen$^{2}$}

\affiliation{
$^1$Department of Chemistry, Ben-Gurion University of the Negev, Beer-Sheva 84105, Israel \\
$^2$Department of Physics, Ben-Gurion University of the Negev, Beer-Sheva 84105, Israel
}

\begin{abstract}
We introduce a semiclassical theory for strong localization 
that may arise in the context of many-body thermalization.  
As a minimal model for thermalization we consider a few-site 
Bose-Hubbard model consisting of two weakly 
interacting subsystems that can exchange particles. 
The occupation of a subsystem ($x$) satisfies in the classical treatment 
a Fokker-Planck equation with a diffusion coefficient $D(x)$. 
We demonstrate that it is possible to deduce from 
the classical description a quantum breaktime $t^*$, 
and hence the manifestations of a strong localization effect.        
For this purpose it is essential to take the geometry 
of the energy shell into account, and to make a distinction 
between different notions of phase-space exploration. 
\end{abstract}
\maketitle


\section{Introduction}

Equilibration in isolated bipartite systems is a major theme in many-body statistical mechanics. Hamiltonian classical or quantum dynamics can emulate thermalization between weakly-coupled constituent subsystems, provided at least one of them is classically chaotic, resulting in an ergodic evolution.  The classical thermalization is then described by a Fokker-Planck equation (FPE) \cite{trm,tmn} depicting a diffusive redistribution between the accessible states,  with an implied fluctuation-dissipation theorem.

While chaos can provide the required ergodicity in a classical thermalization process, the corresponding quantum mechanical thermalization scenario \cite{Srednicki94,Rigol08,Polkovnikov11,Gring12,Olsh2,Basko,Santos,SantosRev} is endangered by the emergence of quantum localization \cite{Anderson,Scaling,QKRc,QKRf,QKRh,QKRb,Smilansky}. Much effort has been invested in the study of many-body localization of large disordered arrays \cite{Mirlin05,Basko06,gora,Berkovits,BarLev,lea1,lea2}, but the physics of localization in such large systems remain ambiguous.
Even the definitions are vague, and the role of semiclassical phase-space structures with regard  to the determination of the mobility edge has not been addressed \cite{BarLevRev}. 
It is therefore essential to consider tractable minimal models for thermalization, in which the origins of localization can be traced. Such models should include two weakly coupled subsystems, where the classical chaoticity requirement imposes a  minimum of two degrees of freedom on at least one of them.

\begin{figure*}

\includegraphics[width=13cm]{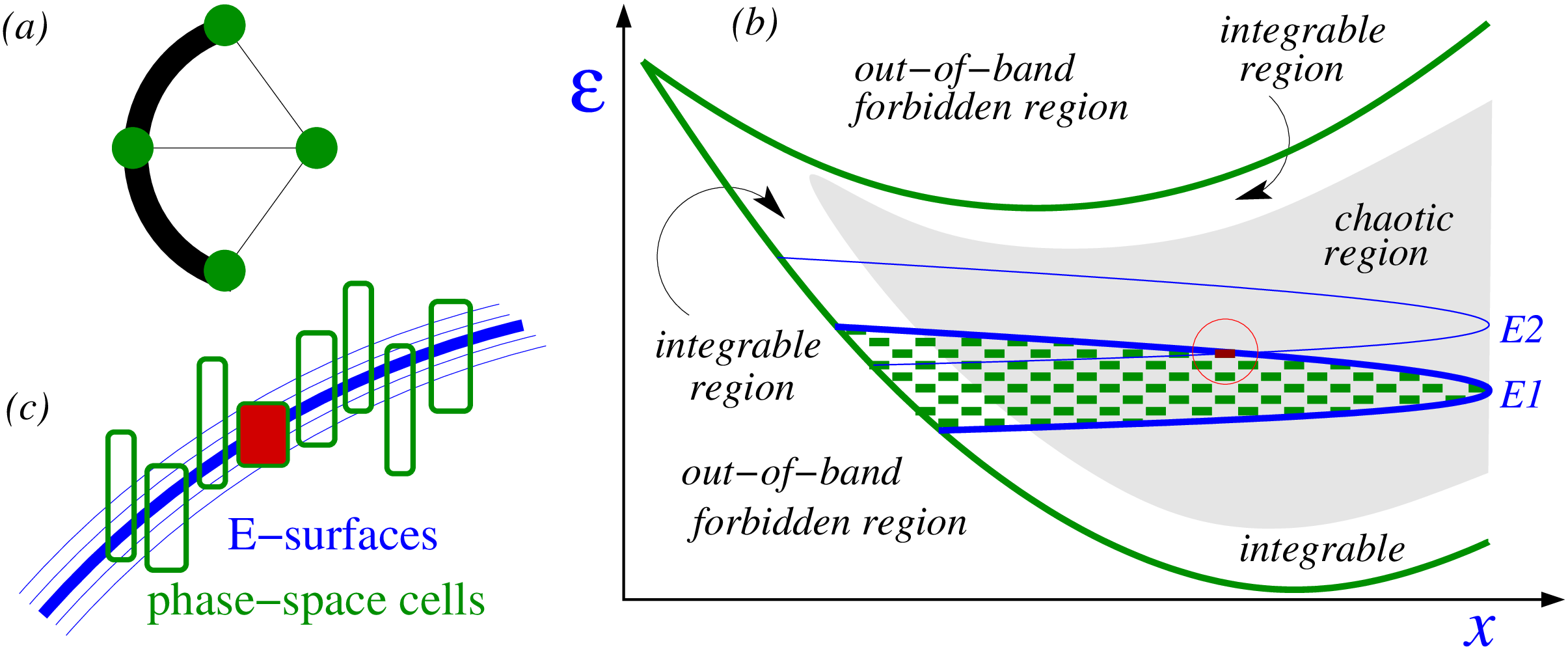}

%
%
\caption{\label{f1}
{\bf Illustration of the system and its phase-space.} 		
{\bf (a)} We consider $N=60$ bosons in a four-site system that is formed by weakly coupling a trimer and a monomer subsystems.  
The trimer consists of three strongly coupled sites. 
The system is described by the Hamiltonian of \Eq{eBHH}.
{\bf (b)}~The phase-space of the system is divided into cells that are labeled by $r=(x,\varepsilon)$.      
A given cell $r_0$ (circled and colored in red) overlaps with  energy surfaces ${E_1 < E < E_2}$, 
forming a region that we call its ``energy shell" (strictly speaking, we display the {\em projection} 
of a high-dimensional energy shell onto the two-dimensional plane).   
Each surface $E$ overlaps with many cells, as shown schematically for $E_1$ 
(the white spaces between the cells are for visual purposes only). 
Note that the width in $\varepsilon$ of the energy surface shrinks to zero for ${x\rightarrow N}$, where the trimer-monomer coupling term \Eq{eBHHp} vanishes.
A single classical trajectory explores a zero-thickness energy surface~$E$, 
and can visit at most $\Omega_E$ cells. To be precise, only a fraction $\mathcal{F}^{\text{cl}}$ 
is explored, because $\Omega_E$ counts not only cells that belong to the chaotic sea, 
but also cells that reside in quasi-integrable regions.
A semiclassical cloud that starts at $r_0$ explores a larger volume 
that includes all the accessible cells within the finite-thickness energy shell. 
{\bf (c)}~An abstract illustration of the high-dimensional energy surfaces. 
Each surface is associated with a quantum eigenstate $E_{\alpha}$.  
The number of surfaces $\mathcal{N}_E$ that participate in the dynamics 
(overlaping with the red cell) might be much smaller than the number 
of cells $\Omega_E$ that intersect a given energy surface.      
} 
\end{figure*}

One type of a model that is experimentally viable \cite{BHH1,BHH2}, is the few-site Bose-Hubbard system \cite{Kottos,sfc,sfa,Henn}. Since the $N$-boson system has a clear classical limit with $1/N$ serving as an effective Planck constant, it is ideal for exploring many-body localization effects in a controlled manner. The two-site Bose-Hubbard system, also known as the bosonic Josephson junction, is excluded due to the integrability of its classical phase-space. The three-site system features low-dimensional chaos \cite{Kottos,sfc}, but it is of little interest for quantum localization studies since its classical phase-space is divided into disjoint territories by Kolmogorov-Arnold-Moser (KAM) tori. The nature of localization in the three-site system is therefore always semiclassical: due to trapping either on a quasi-integrable island, or inside a chaotic pond \cite{sfc}. We therefore conclude that the smallest bi-partite Bose-Hubbard model that may demonstrate a quantum localization effect in its thermalization, is a four-site system \cite{tmn} where the pertinent weakly-coupled subsystems are a chaotic trimer, and a single auxiliary site denoted here as a `monomer'; see \Fig{f1}a.

Preliminary numerical evidence for localization in the dynamics of the four-site model has been obtained in~\cite{tmn}. In some phase-space regions localization is semiclassical: it originates from quasi-integrability and therefore persists in the classical limit. However, there are other phase-space regions that are classically completely chaotic, yet exhibit localization quantum mechanically. This Anderson-type localization does not survive in the classical limit. 

Surprisingly, no semiclassical theory for strong localization in such a minimal model is currently available.
The original view of Anderson \cite{Anderson} holds that strong localization appears due to interference of trajectories. This leads to the Anderson criterion which involves the {\em connectivity} of space. In certain cases it is possible to carry out a semiclassical summation, to identify families of destructively interfering trajectories; see, for example, Ref.~\cite{Smilansky}. However, this approach is a dead-end as far as physical insight is concerned. 
A~different paradigm, namely, the scaling theory of localization~\cite{Scaling}, illuminates the importance of {\em dimensionality}. But, clearly, such an approach is designed for scalable disordered systems, and not for our model of interest which contains finite subsystems with few freedoms,  and where idealized chaos of the random-matrix-theory-type cannot be assumed. We would like to have a theory that will deduce quantum localization from semiclassical simulations, without having to take the details of interference into account. 

In this paper we argue that such a theory can be attained by an extension of a neglected paradigm \cite{brk1,brk2,brk3,brk4} that regards quantum localization as the breakdown of quantum-classical correspondence (QCC). The idea is to figure out a procedure that allows for the semiclassical determination of a quantum breaktime. Such an approach has been discussed in the past with regard to localization in Anderson-type models in $d=1,2,3$ dimensions \cite{brk3,brk4}, but its adaptation for the analysis of localization in complex systems has not been explored. 
Here, we construct the necessary  semiclassical framework for a detailed study of strong quantum localization and present the necessary tools for its analysis. We are inspired by the work of Heller regarding phase-space exploration \cite{Heller,hlc} that has been used in the past mainly in the context of weak localization, a.k.a. scar theory \cite{scar1,scar2}.
We use the four-site Bose-Hubbard model to benchmark this theory and demonstrate its feasibility.

Let us first construct a naive theory. Let $x$ be a coordinate that describes the thermalization process. In our four-site minimal model it is the occupation of the trimer subsystem, with the monomer subsystem containing the remaining $N-x$ particles.
We assume that in the classical description the system is chaotic within the relevant energy range, and accordingly, we can derive an FPE for the evolving probability distribution $p(x;t)$, as explained in \cite{trm}. This FPE requires the calculation of a diffusion coefficient $D(x)$. Inspired by the literature on Anderson localization in quasi-one-dimensional arrays \cite{brk1,brk2,brk3,brk4} we might deduce an emergent localization length $\xi = g(x) D(x)$, where $g(x)$ is the density of states (given~$x$) at the region of interest. It turns out that such an approach does not work. In fact, it should be obvious in advance that it cannot be a generally valid procedure, because the actual dimensionality of the system is completely ignored. Were it valid, it would {have} implied that any diffusing coordinate is doomed to be localized in the quantum treatment, irrespective of the existence of extra coordinates. 

We therefore have to trace back one step, and to recall the argument that leads to the semiclassical expression for ${\xi}$. The idea is to generalize the QCC condition ${t<t_H(\Omega)}$, where $\Omega$ indicates the volume of the system, and $t_H=2\pi/\Delta_0$ is the Heisenberg time, determined by the mean level spacing $\Delta_0$. 
This generalization is performed by replacing the total $\Omega$ by the classically {\em explored} volume~$\Omega^{\text{cl}}_t$, such that the {\em running} Heisenberg time is now related to the {\em effective} level spacing; hence, the QCC condition becomes ${t<t_H(\Omega_t})$. The breakdown of this condition \cite{brk1,brk2,brk3,brk4} determines the breaktime $t^*$, and hence the localization volume.  

The above is roughly the approach we are going to employ. The challenge is to provide a proper phase-space formulation of the QCC condition, taking the non-trivial geometry of the energy shell into account. It is important to realize that the classical {\em exploration} volume, contrary to the intuitive thinking, is not the same volume over which the probability distribution $p(x;t)$ spreads after time $t$, henceforth named the {\em spreading} volume.

\section{Outline} 
We define a quantum localization measure~$\mathcal{F}^{\text{s}}$ and demonstrate the manifestation of strong localization in our model system. The objective is to provide a semiclassical theory for the breaktime. This goal is attained in two stages: (a) The first step is to introduce  definitions for the classical phase-space exploration function $\Omega^\text{cl}_t$, and for the quantum Hilbert-space exploration function $\mathcal{N}^{\text{qm}}_t$. Associated with it is the distinction between $\Omega_E$ that counts phase-space cells that intersect a given energy surface, and $\mathcal{N}_E$ that measures the width of the energy shell.  
This leads naturally to the definition of the classical and the quantum ergodicity measures $\mathcal{F}^\text{cl}$ and $\mathcal{F}^{\text{qm}}$; (b) The second step is to formulate a phase-space version for the QCC condition:
\be{QCC}
\mathcal{N}^{\text{sc}}_t \ \ < \ \ \mathcal{F}^{\text{qm}}_\text{erg}\left[\frac{\mathcal{N}_E}{\Omega_E}\right] \Omega^{\text{cl}}_t~.
\eeq
Here $\mathcal{N}^\text{sc}_t \approx t/t_E$ is the semiclassical estimate for the quantum exploration which depends crucially on a time scale $t_E$, determined by the width of the energy shell. The quantum factor $\mathcal{F}^{\text{qm}}_\text{erg}=1/3$ is required to account for universal quantum fluctuations.

We then demonstrate that the above phase-space version of the QCC condition can be applied in our minimal model for many-body thermalization. It provides a reliable and accurate estimate for the breaktime; from the latter we deduce the localization volume, obtain an estimate for the localization measure, and determine the phase-space mobility edge.

The main text takes the reader in the shortest possible way to section~IX, 
where the semiclassical prediction is compared with the actual quantum results. 
Very important technical issues have been deferred into the appendices. 
In Appendix~A we discuss the notion of an {\em improper} Planck cell. 
In Appendix~B we provide the numerical details of the simulations. 
In Appendix~C we clarify how the phase-space exploration, 
the survival probability and the local density of states are related, 
pointing out a subtle twist regarding the notion of ``semiclassical approximation".

Finally, in Appendix~D we emphasize  that for any high-dimensional chaotic system (namely, with more than two degrees of freedom) the energy shell cannot be divided into separate territories (chaotic sea and islands). Hence  classical localization is strictly-speaking impossible due to a very slow Arnold diffusion process. Consequently, the semiclassical quantum-breaktime perspective is, in fact, formally essential for the discussion of long-time localization, not only in the chaotic sea, but also in quasi-integrable or mixed regions of phase-space.

\section{Model System}

Our benchmark system is the $N$-particles trimer-monomer model, illustrated in \Fig{f1}a.  
Below, the time units are chosen such that ${\hbar=1}$.
The system is described by the Bose-Hubbard Hamiltonian (BHH),
\be{BHH}
\label{BHH}
\mathcal{H} \ = \ \mathcal{H}_0+\mathcal{H}_c~. 
\eeq
The first term on the r.h.s. of Eq.~(\ref{BHH}) is the Hamiltonian of the decoupled subsystems,
\be{BHHz}
\mathcal{H}_0 \ \ = \ \ \frac{U}{2}\sum_{j=0}^{3} \hat{n}_j^2 
-\frac{K}{2} (\hat{a}_3^\dagger\hat{a}_2 +\hat{a}_2^\dagger\hat{a}_1 + \text{h.c.})~.
\eeq
The operators $\hat{a}_j^\dagger$, $\hat{a}_j$ and $\hat{n}_j=\hat{a}_j^\dagger\hat{a}_j$ create, 
destroy and count particles at site~$j$. The $j=0$ site is the monomer, 
while the $j=1,2,3$ trimer-sites form a chain, with hopping frequency~$K$. The parameter $U$ is the on-site interaction strength per particle.
All trimer sites are weakly coupled with hopping frequency ${K_c \ll K,UN}$ to the monomer. 
Accordingly, the monomer-trimer coupling term in Eq.~(\ref{BHH})  is 
\be{BHHp}
\mathcal{H}_c \ \ = \ \  - \frac{K_c}{2} \sum_{j=1}^3 (\hat{a}_0^\dagger \hat{a}_j + \text{h.c.})~.
\eeq

In the absence of coupling the Hamiltonian $\mathcal{H}_0$ conserves the total trimer population 
\beq
\hat{x} \ \equiv \ \hat{n}_1+\hat{n}_2+\hat{n}_3~,
\eeq
and hence $x$ is a good quantum number for the unperturbed eigenstates. Another good quantum number is the scaled energy ${\varepsilon = \langle \mathcal{H}_0\rangle/(NK)}$, hence the eigenstates of the coupling-free system can be denoted as $\ket{r}=\ket{x,\varepsilon}$. We use the same scaling for the perturbed energies $E_\alpha=\langle \mathcal{H}\rangle/(NK)$  associated with the exact eigenstates $\ket{\alpha}$, for which $x$ is no longer a good quantum number.

The classical limit is obtained by replacing the bosonic operators with $c$ numbers, namely $\hat{a}_j\rightarrow \sqrt{n_j}\exp(i\varphi_j)$. Since the Hamiltonian is $U(1)$ invariant, the overall phase is insignificant. Thus, the classical phase-space of the four-site system is six-dimensional, spanned by three pairs of conjugate variables, e.g., the site-population differences $q_j=n_j-n_0$ and the relative phases $p_j=\varphi_j-\varphi_0$, where ${j=1,2,3}$.
 Further reduction into a four-dimensional classical phase-space is possible when there is no coupling. Standard rescaling implies that the dimensionless classical parameters are 
\beq
u=\frac{NU}{K}; \ \ \ \ \ k=\frac{K_c}{K} 
\eeq
while the effective Planck constant is $1/N$.

The classical phase-space of the system may be divided into Planck cells of volume $h^d$, where $d$ is the number of freedoms ($d{=}3$ for our model system), and $h$ is the Planck constant. Technical details and discussion of some subtleties of this partition are provided in \Ap{ap:phase-space_partition}. It is important to emphasize that~$h$ is implicit in the semiclassical context via the definition of the volume of a Planck cell, but otherwise it has no effect on the classical dynamics. In the absence of coupling, each cell consists of all trimer phase-space configurations that have the same ${x}$ (within unit uncertainty) and the same $\varepsilon$ (within level spacing uncertainty, considering the spectrum of a trimer with $x$ particles).
We thus use $r=(x,\varepsilon)$ as a running index to label phase-space cells of the unperturbed system. Each classical cell supports a single quantum eigenstate $\ket{r}$ of the coupling-free system.

A schematic representation of the state-space is presented in \Fig{f1}b. 
The spectrum of the coupling-free system at any given $x$ equals the 
spectrum of an $x$-particle trimer plus a shift due to the monomer energy $(1/2)U (N{-}x)^2$.
Thus, the allowed range of $\varepsilon$ increases with~$x$, resulting in the energetically accessible region marked in the figure.  
Within this allowed region, we distinguish between chaotic and integrable domains. 
At low $x$, the trimer nonlinearity, quantified by $Ux/K$, is too small to generate chaos. 
At higher $x$ values, the central part of the spectrum become chaotic, while both the upper and the lower parts remain regular. 
That is because at the highest trimer energies we have quasi-integrable self-trapped motion, 
whereas at low energies the nonlinear interaction is negligible, resulting in quasi-integrable Rabi-Josephson oscillations. 
For a given~$x$ we use the quantum level spacing statistics of the unperturbed spectrum to detect the chaotic (gray) domain, 
and confirm the result by inspecting classical Poincare sections.

\section{Equilibration and Localization}

The weak coupling $\mathcal{H}_c$ between the constituent subsystems generates transitions 
between the unperturbed quantum eigenstates $\ket{r}$. In the semiclassical perspective 
these are transitions between the Planck cells, which are indicated by rectangles in \Fig{f1}b.  
Starting with a cloud of points at a cell~$r_0$ (marked red), the dynamics, if it is fully chaotic,  
can lead to an ergodic distribution within the energy shell of the full Hamiltonian~$\mathcal{H}$. 
In such a case we regard the dynamical process as ``equilibration".   
Projecting this six-dimensional energy shell onto the $(x,\varepsilon)$ plane we get a thin strip.
Assuming that the classical points within the $r_0=(x_0,\varepsilon_0)$ cell, have energies $E \in [E_1,E_2]$, 
the strip can be regarded as the union of partially overlapping sub-strips, each of them 
is the projection of a mono-energetic energy {\em surface}~$E$.       

We describe the quantum and the semiclassical dynamics in $r$ space on an equal footing. For this purpose we define the distribution of probability $P_t(r|r_0)$ to find the system in~$r$ after time~$t$, given that it was launched initially at~$r_0$. In the semiclassical case, this distribution is the fraction of cloud-points that occupy the cell~$r$ at time~$t$, whereas in the quantum case, 
\be{quantum}
P^{\text{qm}}_{t}(r|r_0) \ \ = \ \ \Big| \BraKet{r}{\, \exp(-i\mathcal{H}t)\, }{r_0} \Big|^2~.
\eeq
Both quantum-mechanically and semiclassically, the saturation profile that is obtained 
in the limit $t\to\infty$ can be calculated via a convolution, namely,  
\be{SatProfile}
P_{\infty}(r|r_0) \ \ = \ \  \sum_{\alpha} \rho(r|{E}_\alpha) \, \rho(r_0|{E}_\alpha)~.
\eeq
In the quantum case, the values
\be{LDOS}
\rho(r|E_{\alpha}) \ \ = \ \ \left| \braket{r| E_{\alpha}} \right|^2  
\eeq
are the overlaps between the exact eigenstates $\ket{E_{\alpha}}$ of $\mathcal{H}$, 
and the unperturbed eigenstates $\ket{r}$ of $\mathcal{H}_0$. 
In the semiclassical case, they are the overlaps between energy shells in phase-space;  
see \Ap{ap:phase-space_partition}. 
It should be noticed that, by definition, \Eq{eSatProfile} represents the infinite-time-average of the probability distribution. In a semiclassical simulation $P_t(r|r_0)$ coincides with $P_{\infty}(r|r_0)$ provided~$t$ is much larger than the ergodic time. But in a quantum simulation, fluctuations and rare recurrences persist for any~$t$, and therefore the time averaging becomes essential for the definition of a saturation profile.

\Eq{eSatProfile} has been numerically verified by a long-time numerical quantum propagation.
Saturation distributions are shown in \Fig{f2} for a semiclassical simulation (panel a) and a quantum simulation (panel b), 
launched at the same~$r_0$. Comparison of the two distributions indicates localization in the quantum case. 
In panel \ref{f2}c, the two distributions are projected onto the $x$-axis to give the $x$-distribution,
\beq
P_{\infty}(x|x_0)=\sum_{r\in x}  P_\infty(r|r_0)~,
\eeq
and compared with the density of states~$g(x)$. 
While the semiclassical distribution has clearly ergodized, 
i.e., $P_{\infty}^{\text{cl}}(x|x_0) \propto g(x)$, 
the quantum distribution remains localized in the large-$x$ region.

\section{Dynamical localization measure}

\begin{figure}
\centering 
\includegraphics[width=\hsize]{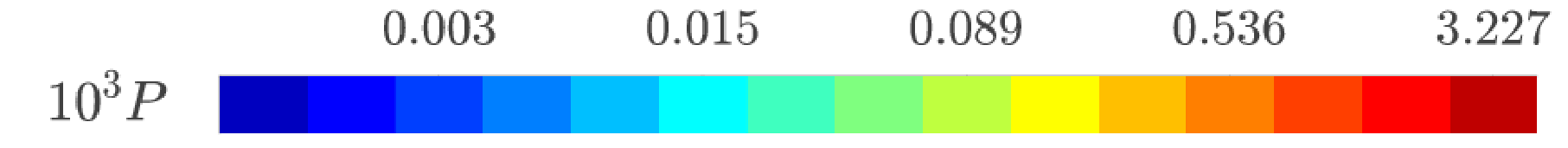}\\
\includegraphics[width=\hsize]{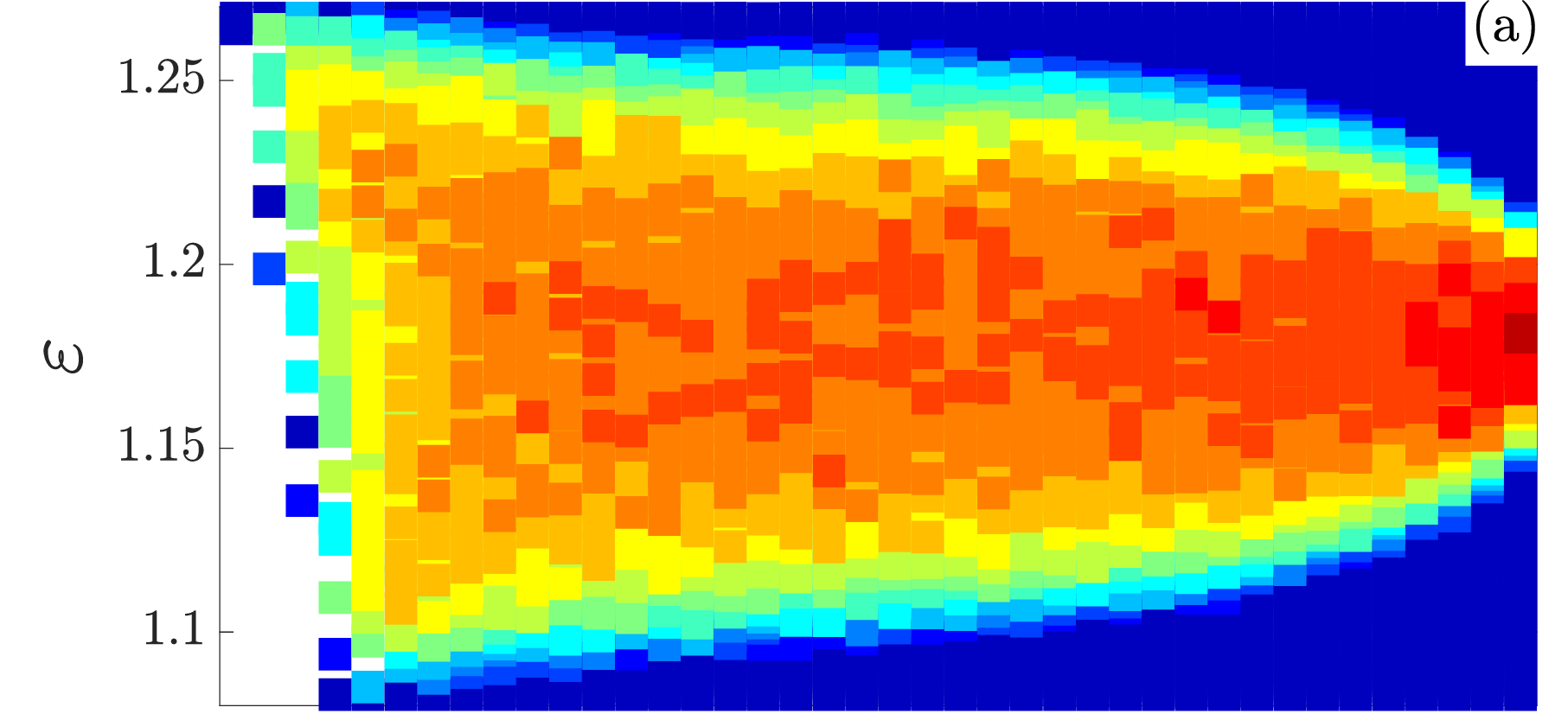}\\
\includegraphics[width=\hsize]{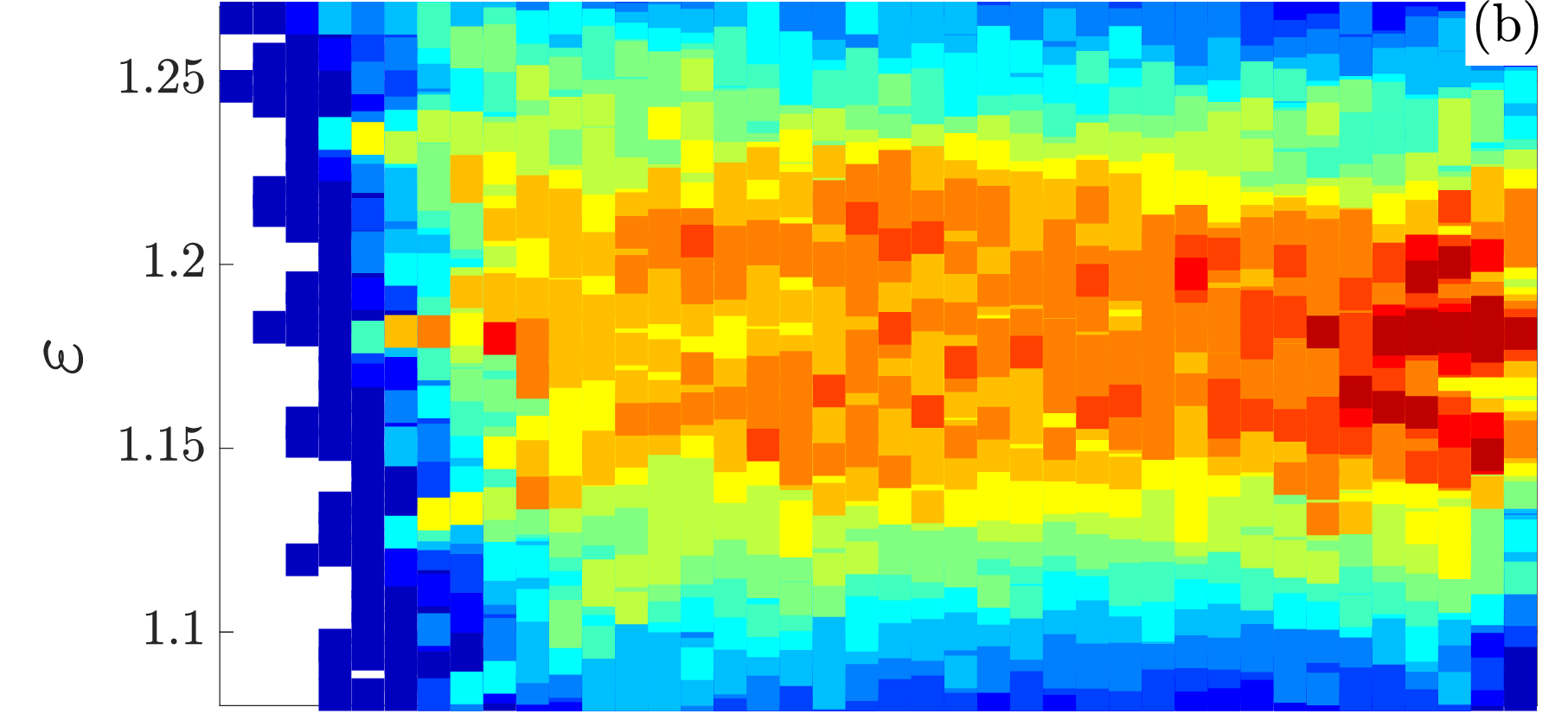}\\
\includegraphics[width=\hsize]{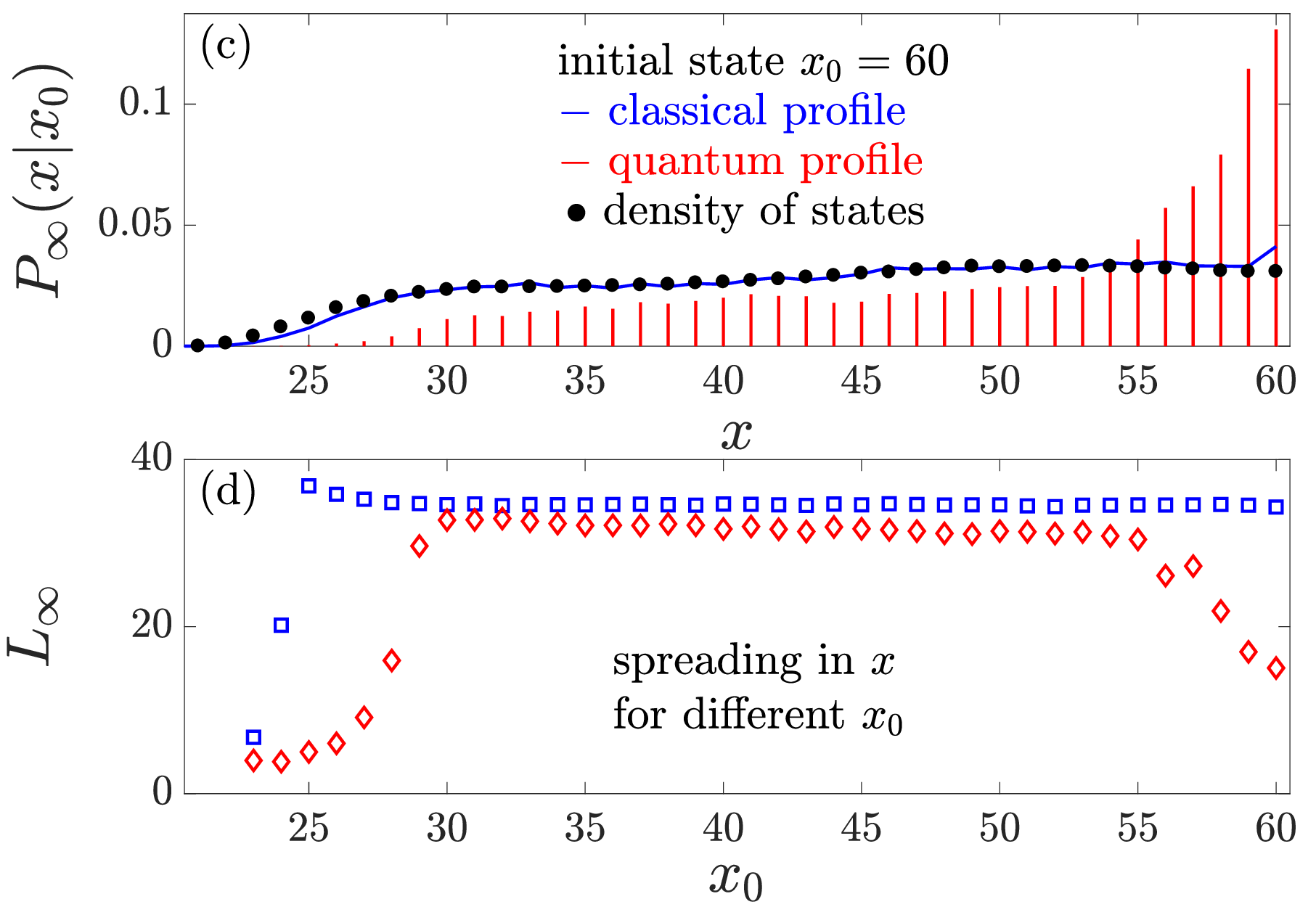}
\caption{\label{f2} 
{\bf The quantum localization effect.} 
Panels~(a) and~(b) display the two-dimensional saturation profile $P_{\infty}(r|r_0)$ for the semiclassical and the quantum simulations, respectively. The initial condition $r_0$ corresponds to having all particles in the trimer ($x_0{=}60$) with  ${\varepsilon_0=1.181}$. The nonlinear color scale encodes the probability from low (blue) to high (red). 
The semiclassical simulation in (a) provides the determination of the dynamically
accessible volume of the energy shell.  
The quantum simulation exhibits strong localization. In panel~(c) the distribution is projected 
onto $x$ space. The quantum (red) simulation features a peak at the initial~$x_0$, 
unlike the semiclassical (blue) simulation that reflects phase-space ergodicity, 
as implied by the agreement with the (black) normalized density of states. 
Panel (d) displays the spreading length $L_{\infty}$ for different values of $x_0$, 
at the same $\varepsilon_0$ as in panels~(a-c). 
Note that for small $x_0$ the energy shell gets wider, and hence $L_\infty$ becomes higher.
The abnormally low values of the semiclassical spreading (blue symbols) for $x_0\leq24$ indicate a 
lack of classical ergodicity, and hence are of no interest for us. 
Our objective is to provide a theory for the {\em quantum} spreading (red symbols), 
where strong localization shows up in the range ${25\leq x_0 \leq 29}$ and for ${x_0\geq 56}$. Details of the simulations can be found in \Ap{ap:details}.
}
\end{figure}

Given an initial state $r_0$, it is possible to define a spreading volume $\Omega_t$ that counts how many $r$-locations participate in the quantum or semiclassical $P_t(r|r_0)$ distribution:
\be{Omega}
\Omega^{\text{qm/sc}}_t \ \ = \ \ \left\{ \sum_r \Big[ P_t(r|r_0) \Big]^2 \right\}^{-1}~.
\eeq
The semiclassical saturation value $\Omega^\text{sc}_\infty$ 
reflects the dynamically accessible volume of the energy shell. 
Similarly, one can define for the same initial preparation the spreading volume $L_t$ in $x$,
\be{Lx}
L_t \ \ = \ \ \left\{ \sum_x \Big[ P_t(x|x_0) \Big]^2 \right\}^{-1}~,
\eeq
whose saturation value $L_\infty$ reflects the accessible $x$ volume.
Results for the saturation value $L_{\infty}$, as a function of the initial value $x_0$, are displayed in \Fig{f2}d.

Dynamical localization is implied if the accessible quantum spreading volume is less than the corresponding semiclassical spreading volume. We thus define a dynamical localization measure as the fraction  
\be{FDL}
\mathcal{F}^{\text{s}} \ \ \equiv \ \ 
\frac{\Omega^{\text{qm}}_\infty}{\Omega^{\text{sc}}_\infty}~.
\eeq 
Strong dynamical localization means that the quantum distributions occupies only a small fraction of the semiclassical spreading volume, and hence ${\mathcal{F}^{\text{s}} \ll 1}$. 

To clarify the semantics, we should at this point discuss the relation 
between the disorder-induced {\em Anderson localization} and the chaos-induced {\em dynamical localization}.  
In the Anderson model $\Omega^\text{sc}_\infty$ corresponds to the total volume $L^d$ 
of a $d$-dimensional disordered lattice, while $\Omega^{\text{qm}}_\infty$ corresponds 
to some localization volume~$\xi^d$. The term ``dynamical localization" has been introduced 
in the quantum chaos literature in connection with the standard map, a.k.a. the ``kicked rotor", 
where the explored ``locations" are angular momentum states \cite{QKRc,QKRf,QKRh,QKRb}. 
By now it is recognized that both {\em disorder} and {\em chaos} can lead to strong localization effect. 
In the perspective of the present work, both are implied by a breakdown of a QCC condition, 
and can be handled on equal footing.

\section{State-space exploration}

To follow our formulation below, it is crucial to distinguish between state-space {\em spreading}, referring 
to the instantaneous fraction of space occupied by a time dependent distribution at time $t$, as described in the previous section,
and {\em exploration}, referring to the accumulated state-space `volume' visited during time~$t$.
 
The notion of exploration is treated on an equal footing for both the classical and the quantum cases. The classical definition 
of ``phase-space exploration" is inspired by past work on lattice random walk  \cite{Montroll}, 
whereas the quantum notion of ``Hilbert-space exploration" is adopted from \cite{Heller}. 
The evolving state of a system is described by a delta-distribution (a point) in phase-space 
in the classical case (``cl"), or by a cloud of points in the semiclassical case (``sc"), 
or by a probability matrix in the quantum case (``qm"). In all three cases, the instantaneous state of the system 
is denoted as $\varrho(t)$, with $\overline{\varrho}(t)$ being its average during the time interval~$[0,t]$,
\beq
\overline{\varrho}(t) 
\ \ \equiv \ \
\frac{1}{t}\int_0^t \varrho(t')dt' ~.
\eeq
The explored space is then defined as 
\be{TraceSquare}
\left\{ \amatrix{ \Omega^\text{{cl}}_t \cr \mathcal{N}^\text{{qm}}_t  } \right\}
\ \ \equiv \ \  
\left\{ \trc\left[ \overline{\varrho}(t)^2 \right] \right\}^{-1}~.
\eeq
The classical (quantum) function  $\Omega^\text{{cl}}_t$ ($\mathcal{N}^\text{{qm}}_t$) 
provide the minimal number of classical phase-space cells (quantum basis-states) 
required to describe the time-dependent dynamics up to time~$t$. 
Namely, the classical function $\Omega^\text{{cl}}_t$ counts how many cells have been visited by a classical trajectory, while the quantum function $\mathcal{N}^\text{{qm}}_t$ counts the number of states that  have participated in the dynamics during this time.

As observed in \cite{Heller}, the number of explored quantum states  $\mathcal{N}^\text{{qm}}_t$ is related to the survival probability $\mathcal{P}(t)$ of the initial state $|r_0\rangle$, and the latter is related via a Fourier transform to the the local density of states (LDOS), which we denote as $\rho(E)$. 
We summarize the precise definitions, and critically clarify these relations in Appendix~C.  
For the subsequent presentation the main points are as follows:
{\bf (1)}~Given an initial preparation, the associated semiclassical LDOS 
can be used in order to define the width~$\Delta_E$ and the volume~$\mathcal{N}_E$ 
of the energy shell.  
{\bf (2)}~The quantum LDOS provides the definition of~$\mathcal{N}_{\infty}$, which is the number 
of energy eigenstates that actually participate in the time evolution.
{\bf (3)}~The semiclassical approximation $\mathcal{N}^{\text{sc}}_t$ cannot be defined via 
the semiclassical evolution using \Eq{eTraceSquare}, but rather has to be defined 
from the semiclassical LDOS.
{\bf (4)}~The definition of the quantum time step~$t_E=2\pi/\Delta_E$ is implied by 
the semiclassical approximation. It is the smallest time that can 
be resolved by the quantum evolution. It should be contrasted 
with the Heisenberg time time~$t_H=2\pi/\Delta_0$, which is the upper limit 
for the manifestation of quasi-periodicity. Note that $\mathcal{N}_E=t_H/t_E$.

\section{Measures for classical and quantum ergodicity}

Since we consider systems with a mixed phase-space, containing integrable as well as chaotic regions, ergodization is necessarily incomplete. It is, therefore, important to quantify the degree of classical and quantum ergodicity. The classical ergodicity measure is
\be{FCL}
\mathcal{F}^{\text{cl}} \ \ \equiv \ \
\frac{\Omega^{\text{cl}}_\infty}{\Omega_E}~,
\eeq 
where $\Omega_E$ is the total number of cells that overlap a typical energy surface $E$ within the energy shell; see \Fig{f1}b.
This measure reflects the relative volume of the chaotic sea within the energy shell. 
Thus, $\mathcal{F}^{\text{cl}}=1$  is only obtained for a fully chaotic energy shell without any quasi-integrable islands. 
At this point is useful to note that based on the convolution formula \Eq{eSatProfile} 
we expect the relation $\Omega^{\text{sc}}_{\infty} \approx \Omega^{\text{sc}}_E $, where
\be{OSC}
\Omega^{\text{sc}}_E 
\ \ = \ \   \sqrt{  \mathcal{N}_E^2 +  \Omega_E^2 }~,
\eeq
In the above formula $\mathcal{N}_E$, unlike $\Omega_E$, is $r_0$-dependent. 
If the energy shell has a trivial ``flat" geometry, such that $\mathcal{N}_E$ unperturbed states mix into $\mathcal{N}_E$ perturbed states in the same energy range $\Delta_E$, then it follows that ${\mathcal{N}_E = \Omega_E}$, and hence $\Omega^{\text{sc}}_E \approx \sqrt{2}\, \Omega_E$.

The quantum ergodicity measure, as proposed in \cite{Heller}, 
is defined in a way that is analogous to \Eq{eFCL}. 
Here we refer to {\em Hilbert-space} exploration rather than {\em Phase-space} exploration.  
Namely, 
\be{FS}
\mathcal{F}^{\text{qm}} \ \ \equiv \ \ \frac{\mathcal{N}_{\infty}}{\mathcal{N}_E}~.
\eeq 
Unlike $\mathcal{F}^{\text{cl}}$, the ergodic maximal value of $\mathcal{F}^{\text{qm}}$ is not unity. 
For quantum-ergodic dynamics of the GOE type one expects $\mathcal{F}^{\text{qm}}_\text{erg}=1/3$ due to the universal effect of quantum fluctuations, and the statistical nature of the quantum-ergodic distribution (see Eq.36 of \cite{Heller}).

\begin{figure}
\centering 
\includegraphics[width=\hsize]{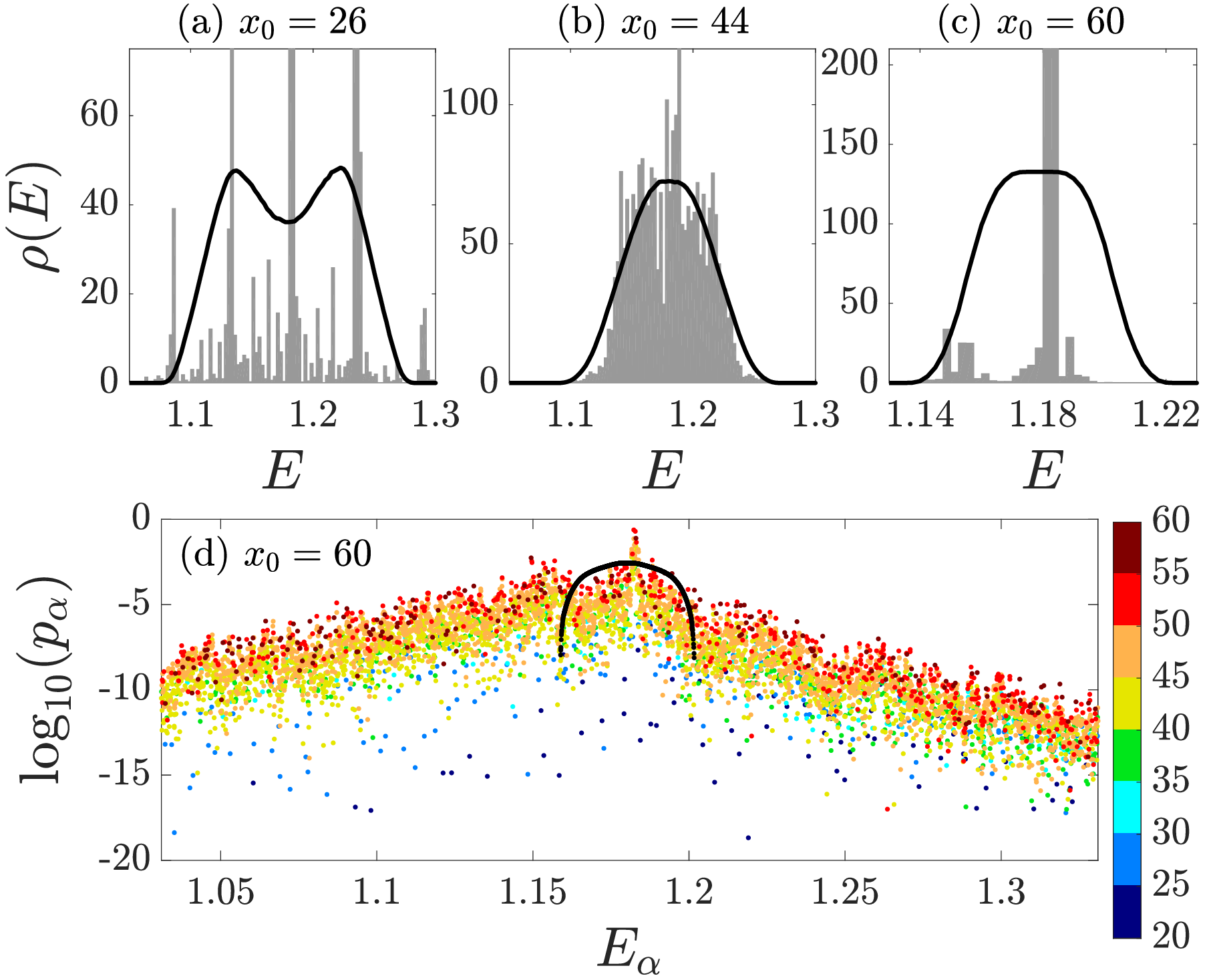}
\caption{\label{f3} 
{\bf Signatures of localization in the LDOS.} 
Panels (a,b,c) compare the quantum LDOS (gray) with its semiclassical counterpart (black).
The calculation is done for the same preparation as in \Fig{f2}, 
that has the energy ${\varepsilon_0 = 1.181}$, and for two other 
preparations with the same energy but with different initial occupations (${x_0=26,44,60}$).    
The resolution has been improved by integrating the density $\rho(E)$ 
over an energy range that corresponds to 50 level spacings.
Quantum ergodicity is reflected in panel (b) where the LDOS matches well the semiclassical envelope. Quantum localization is reflected in panels (a) and (c). The vertical axis has been zoomed in (a,c) and hence the peaks are chopped. Note also the reduced range of the horizontal axis in panel (c).  
Panel (d) provides a sharper view of panel (c). It displays the bare probabilities $p_{\alpha}$ instead of the smoothed density $\rho(E)$, and uses a log scale for the vertical axis. The quantum symbols are color-coded according to the value of $\langle x\rangle_\alpha$. The semiclassical LDOS is the black line. 
We observe that localization is present both in $E$ and in $x$. The localization in $x$ is reflected as sparsity: there are few low-lying blue points that correspond to small $\langle x\rangle_\alpha$ values, and many high-lying red points that correspond to large values.
}
\end{figure}

It should be clear from the illustration in \Fig{f1}c that, in general, the number of quantum states in the energy shell $\mathcal{N}_E$ can be much smaller than the number of Planck cells $\Omega_E$ that intersect a typical energy surface~$E$.  The semiclassical LDOS $\rho(E)$ is the overlap of the initial Planck cell $r_0$ with various energy surfaces of the perturbed system.  Similarly, the quantum LDOS is the projection of the initial state~$|r_0\rangle$ onto the perturbed eigenstates~$\ket{E_{\alpha}}$.  The two distributions are shown in \Fig{f3} for three representative preparations. Dynamical localization is implied when the quantum LDOS does not fill the semiclassical LDOS envelope. This can be simply due to it being narrower than the semiclassical width (localization in $E$) or due to its `sparsity' within the semiclassical envelope (localization in $x$). Either way, the number of  dynamically accessible quantum states becomes much smaller than the number of dynamically accessible classical Planck cells. The system is classically ergodic but quantum mechanically localized.

\begin{figure}
\centering 
\includegraphics[width=\hsize]{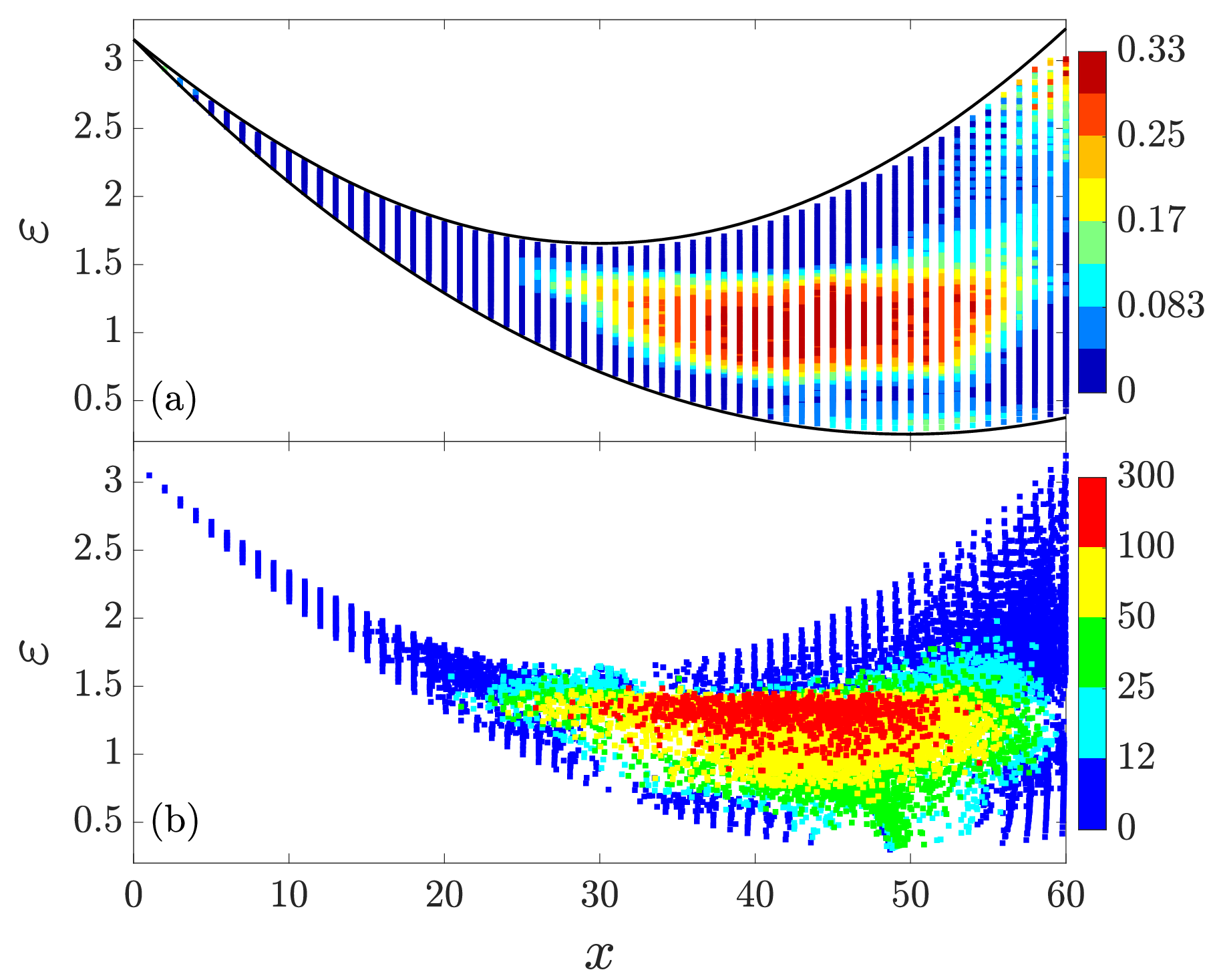}
\caption{\label{f4} 
{\bf The quantum spectrum.} 
{\bf (a)} The unperturbed states $\ket{r}=\ket{x,\varepsilon}$ are 
color-coded according to $\mathcal{F}^{\text{qm}}$. Red color (high $\mathcal{F}^{\text{qm}}$) implies 
quantum ergodicity, as in the LDOS of \Fig{f3}b.
By contrast, blue color (low $\mathcal{F}^{\text{qm}}$) indicates strong quantum localization, as in \Fig{f3}c. 
{\bf (b)} The perturbed states $\ket{E_\alpha}$ are color-coded according to $\text{var}(x)_{\alpha}$, and positioned according to  $(\braket{x}_{\alpha},\braket{\varepsilon}_{\alpha})$. 
The low-variance states (blue) have significant overlaps only with unperturbed states for which $x\approx \langle x\rangle_\alpha$, while the large-variance states (red) correspond to microcanonical states within the chaotic sea.}
\end{figure}

\Fig{f4}a displays the entire spectrum of the possible unperturbed preparations $\ket{r_0}$.
For each $r_0$ the LDOS is calculated, and $\mathcal{F}^{\text{qm}}$ 
is extracted. The red-coded states are quantum-ergodic, 
while the blue-coded states reside in region where the eigenstates 
are localized. This is confirmed by \Fig{f4}b where 
the eigenstates~$\ket{E_{\alpha}}$ are color-coded according 
to their $\text{var}(x)_{\alpha}$.  Thus, the high $\langle x \rangle$ blue points in  \Fig{f4}b
correspond to eigenstates which reside in a chaotic region, but due to their $x$-localization, do not conform to the
Eigenstates thermalization hypothesis \cite{Srednicki94,Rigol08}.

\section{Spreading vs Exploration - Numerical results}

Summarizing the discussion so far, we have distinguished between `spreading' and `exploration' functions. The former ($\Omega_t^{\text{qm}}$ and  $\Omega_t^{\text{sc}}$) count the instantaneous number of $|r\rangle$~states (or $r$ cells) that are occupied by $P_t(r|r_0)$ at time~$t$, while the latter ($\mathcal{N}_t^{\text{qm}}$ and $\Omega_t^{\text{cl}}$) measures the Hilbert-space (or phase-space) dimension that is required to trace the quantum (or classical) dynamics {\em up to time~$t$}.

The initial growth of $\mathcal{N}_t^{\text{qm}}$ is approximated by the semiclassical linear behaviour $\mathcal{N}_t^{\text{sc}}$ of \Eq{ensclin}. It is important to realize that the latter does not reflect the semiclassical evolution: it is not obtained from \Eq{eTraceSquare}, but rather from \Eq{eNSC}. The degree of correspondence between $\mathcal{N}_t^{\text{qm}}$ and $\mathcal{N}_t^{\text{sc}}$ is merely a reflection of the LDOS-correspondence, as was already discussed in \Fig{f3}. To the extent that the semiclassical envelope agrees with the actual quantum envelope, short-time correspondence is guaranteed by definition. In contrast, short-time QCC for the actual time evolution of the spreading is not implied by the LDOS analysis.

In \Fig{f5} we display an example for the time evolution of the `spreading' and `exploration' functions.
The classical exploration is described by $\Omega^{\text{cl}}_t$, that corresponds to the number of cells visited by a {\em single} classical trajectory during the time~$t$. Its growth is much slower compared to the semiclassical spreading $\Omega^{\text{sc}}_t$, which corresponds to the number of cells occupied by a cloud of classical trajectories at time~$t$. 
In fact, $\Omega^{\text{cl}}_t$ does not even saturate during the displayed time interval. However, after a much longer simulation time, it does reach the saturation value $\Omega_{\infty}^{\text{cl}}$ which is also indicated in the figure. The latter value corresponds to the entire chaotic fraction of the energy shell. Similar saturation values are obtained for other initial conditions, as shown in \Fig{f6}a. The near unity value of $\mathcal{F}^{\text{cl}}$ indicates that the system is classically ergodic for all $r_0$ with ${x_0>25}$. We note that the boundary of the chaotic region in \Fig{f1}b has been determined numerically in \cite{tmn} using a different method.

The slowness of the classical exploration constitutes 
an indication for the high dimensionality of the phase-space, 
and plays a major role in the determination of the breaktime.
Now we want to shift our attention to the 
spreading functions $\Omega_t^{\text{qm}}$ and $\Omega_t^{\text{sc}}$ 
that characterize the actual time evolution.  
The questions that should be asked are:
{\bf (i)}~In what sense do we observe QCC, and for what duration of time?  
{\bf (ii)}~Do we observe classical or quantum localization?   
{\bf (iii)}~Can we deduce the quantum dynamics from the classical dynamics? 
We shall address these questions in the following section.

\section{QCC, breaktime and localization}

\begin{figure}
\centering 
\includegraphics[width=\hsize]{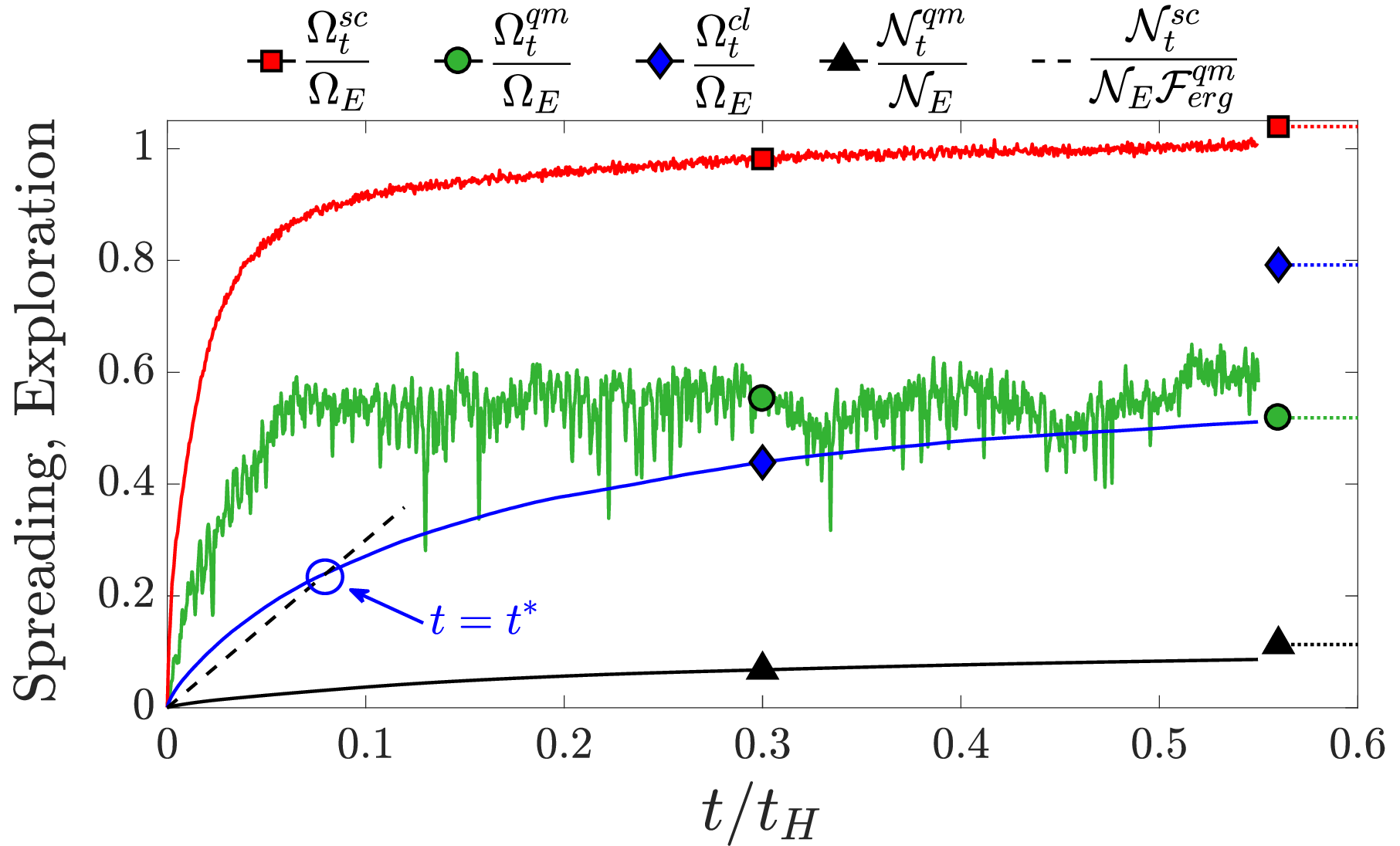}
\caption{\label{f5} 
{\bf Breaktime determination.} 
The functions $\Omega^{\text{cl}}_t$, $\Omega^{\text{sc}}_t$,  $\Omega^{\text{qm}}_t$, and $\mathcal{N}^{\text{qm}}_t$ are plotted 
versus time (see legend) for a preparation that has an initial occupation ${x_0=55}$ with energy ${\varepsilon_0 = 1.181}$.
For further numerical details see Appendix~C. 
The saturation values are indicated by dotted horizontal lines. 
The semiclassical estimate for the breaktime, based on \Eq{eQCC}, 
is determined by the intersection of the dashed line with $\Omega^{\text{cl}}_t/\Omega_E$.}
%
\ \\
%
\centering 
\includegraphics[width=\hsize]{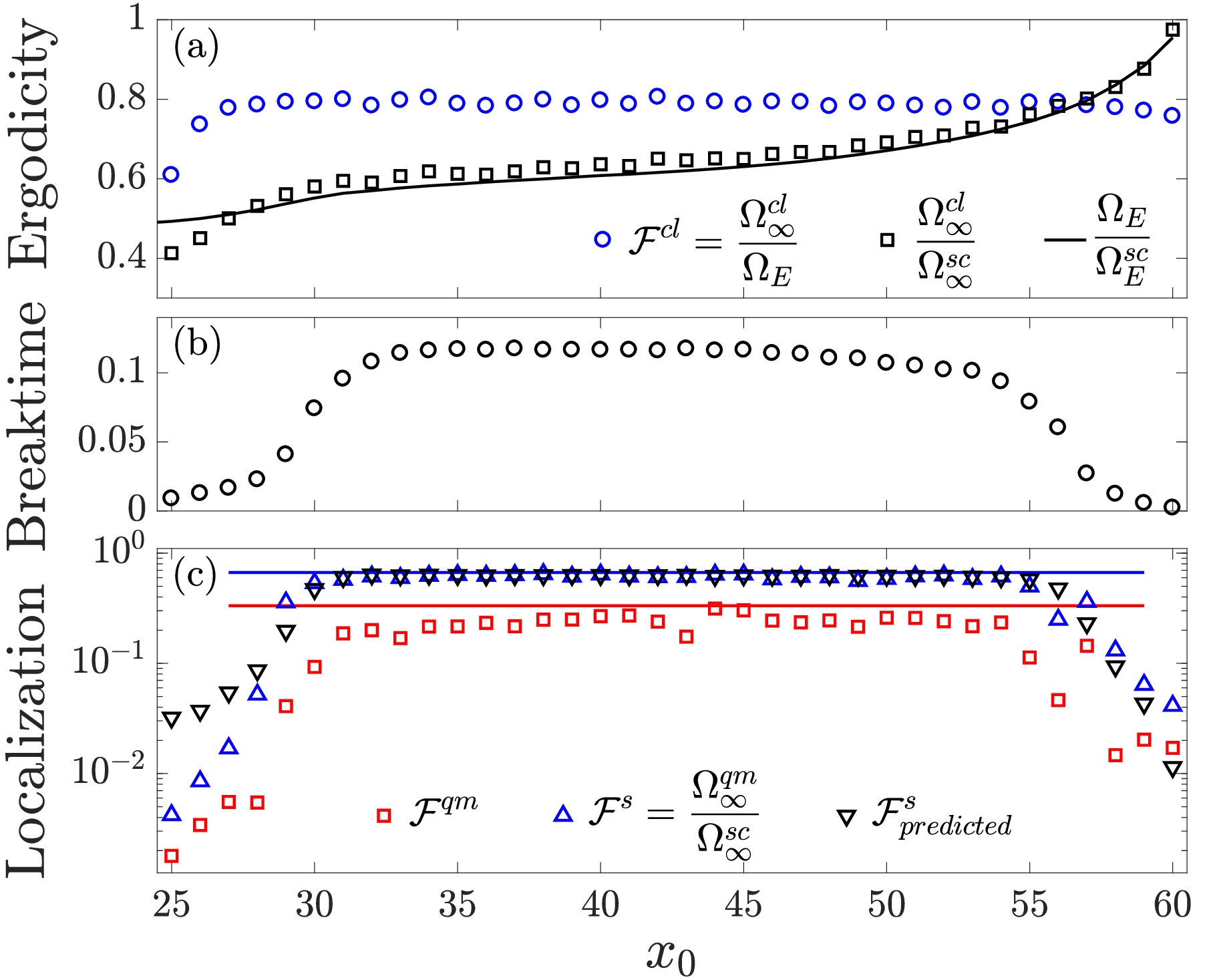}
\caption{\label{f6} 
{\bf Semiclassical prediction of strong localization.}  
The data points in the present figure are based on simulations of the type 
presented in \Fig{f5}, with the same~${\varepsilon_0}$, but for different values of~$x_0$ 
(same simulations as in \Fig{f2}d).
{\bf (a)} The classical ergodicity measure is calculated for each~$x_0$.
Our interest is focused in the range of $x_0$ where $\mathcal{F}^\text{cl}$ indicates a nearly ergodic classical motion. Note that 100\% ergodicity cannot be reached because each energy surface contains inaccessible quasi-regular regions. A~secondary test for ergodicity is the agreement between the exploration-spreading ratio (squares) and it ergodic value (line) which is implied by \Eq{eOSC}. 
{\bf (b)} The scaled breaktime $t^*/t_H$ is deduced from \Eq{eQCC} via the procedure that has been illustrated in \Fig{f5}.
{\bf (c)} 
The quantum ergodization measure $\mathcal{F}^{\text{qm}}$ and the dynamical localization measure $\mathcal{F}^{\text{s}}$ for different initial conditions. 
The horizontal red and blue lines mark the ergodic values $\mathcal{F}^{\text{qm}}_\text{erg}=1/3$ and $\mathcal{F}^{\text{s}}_\text{erg}=2/3$, respectively, that are attained for simulations with ${30\leq x_0 \leq 55}$.   
The prediction for $\mathcal{F}^{\text{s}}$ is based on the semiclassical breaktime estimate \Eq{eLocVolume}.  
The deviation of actual  $\mathcal{F}^{\text{s}}$  from the predicted value at small $x_0$ 
is apparently related to remnants of quasi-integrability.}
\end{figure}

It is generally believed that QCC holds for short time evolution; this assertion is sometimes called ``Ehrenfest theorem". However, this type of QCC is barely relevant in the context of quantized chaotic systems. After an extremely short time scale (the so-called {\em Ehrenfest time}) interference starts to manifest itself, leading to strong quantum fluctuations.  This should be contrasted with the {\em classical mixing} effect that tends to smooth out the time evolution of expectation values. Any meaningful comparison between the quantum and the classical evolutions should treat properly these fluctuations.

Running simulations of the type presented in \Fig{f5}, for initial occupations in the range ${30 \leq x_0 \leq 55}$, we realize in \Fig{f6}c that a maximal value $\mathcal{F}^{\text{s}} \approx 2/3$ is attained. We see in this figure that for the same range the LDOS implies quantum ergodicity with $\mathcal{F}^{\text{qm}}_{\text{erg}} \approx 1/3$. The values~$2/3$ and~$1/3$ are less than a unity due to the quantum fluctuations that have been mentioned in Section~VII. The remaining question is why do we have an extra factor of~${\sim}2$ in the spreading. 
We note that a similar type of factor of~${\sim}2$ has been discussed in the context of the prototype quantum-kicked-rotor problem, see Section~4 of~\cite{brk2}, but the explanation there is not applicable here.
What we have here is an issue with mixed phase space.   
The detailed explanation is provided in Appendix~B2. 
It should be clear that the~${\sim}2/3$ is not universal but has to be determined per-system.

Considering simulations with ${30 \leq x_0 \leq 55}$, we have verified (not displayed) 
that $\Omega_t^{\text{qm}}$ agrees well with $\mathcal{F}^{\text{s}}_{\text{erg}}\Omega_t^{\text{sc}}$. 
Furthermore, we have verified (not displayed) that for the same simulations $L_t^{\text{qm}}$ 
also agrees well with $L_t^{\text{sc}}$. In the latter case the fluctuations are not 
an issue, because the projected distribution $P_t(x|x_0)$ is smooth, unlike $P_t(r|r_0)$.  

Let us look again at the simulation of \Fig{f5}.
Do we have QCC there? Multiplying $\Omega_t^{\text{sc}}$ by $\mathcal{F}^{\text{s}}_{\text{erg}}$ 
we realize that indeed, disregarding fluctuations, a reasonable QCC persists for any time; 
however, for similar simulations with ${x_0>55}$ or ${x_0<30}$, QCC is broken after a short time, 
and we observe that $\Omega_t^{\text{qm}} < \mathcal{F}^{\text{s}}_{\text{erg}}\Omega_t^{\text{sc}}$ 
for later times. This observation is documented in \Fig{f6}c.
We see that for simulations with ${x_0 > 55}$ we get $\mathcal{F}^{\text{s}} \ll 2/3$,  
which implies an earlier saturation for the quantum spreading. 
Similar observation applies for simulations with ${x_0 <30}$.  

We would like to evaluate the time at which the quantum dynamics departs from the classical chaotic dynamics.  
For ballistic motion the QCC breaks down at the Heisenberg time ${t_H=2\pi/\Delta_0}$. For a more general type of dynamics the Heisenberg time is merely an upper bound.
For diffusive systems it has been suggested to define the running Heisenberg time, which is determined by the number of sites that are explored during a random walk process. This corresponds in our phase-space formulation to the explored volume $\Omega^{\text{cl}}_t$. Since~$t_H$ is calculated for the total volume, it follows that the running Heisenberg time is  ${t_H(\Omega^{\text{cl}}_t) = [\Omega^{\text{cl}}_t/\Omega_E]t_H}$. 
Consequently, the QCC condition takes the form ${t < t_H(\Omega^{\text{cl}}_t)}$.

The fastest dynamical time scale is the inverse width of the energy shell $t_E$; see \Eq{edeltaE}. 
The ratio $t_H/t_E$ equals $\mathcal{N}_E$, as implied by \Eq{eNE}.
It follows that the QCC condition can be written as ${ (t/t_E) < [\mathcal{N}_E/\Omega_E]\Omega^{\text{cl}}_t }$. 
We identify the left hand side as the semiclassical approximation $\mathcal{N}^{\text{sc}}_t$ 
for the Hilbert-space exploration function \Eq{ensclin}.  
We also know that for a quantum-ergodic system in a ``flat" fully chaotic energy shell, 
the saturation is attained once ${\mathcal{N}^{\text{qm}}_t \approx \mathcal{F}^{\text{qm}}_{\text{erg}} \mathcal{N}_E}$, 
with  $\mathcal{F}^{\text{qm}}_{\text{erg}}=1/3$ for GOE statistics. 
We therefore conjecture that the general QCC condition is \Eq{eQCC} without any undetermined prefactors.

The breaktime~$t^*$ is the time at which the QCC condition \Eq{eQCC} breaks down. 
Its determination for our model system is carried out by looking for 
the intersection of two {\em classically calculated} curves, 
namely, $\Omega^{\text{cl}}_t/\Omega_E$ and $\mathcal{N}^{\text{sc}}_t/(\mathcal{N}_E \mathcal{F}^{\text{qm}}_{\text{erg}})$,  
as illustrated in \Fig{f5}. Disregarding the Planck-cell partitioning of phase-space, 
no quantum ``input" is required for this procedure.  
The results for other values of~$x_0$ are presented in \Fig{f6}b.

Having found the breaktime, the quantum saturation volume is estimates as follows: 
\be{LocVolume}
\Omega^\text{qm}_\infty \Big|_{\text{predicted}} 
\ \  = \ \ \mathcal{F}^{\text{s}}_{\text{erg}} \ \Omega^\text{sc}_{t*}~.
\eeq
The ergodic value $\mathcal{F}^{\text{s}}_{\text{erg}} \approx 2/3$ 
is used here as a calibration factor. 
Thus, using \Eq{eLocVolume} we obtain a prediction for the localization measure  $\mathcal{F}^{\text{s}}$ 
for any other value of~$x_0$. The results are summarized in \Fig{f6}c, 
and the agreement with the quantum simulation is surprisingly good.

For completeness, we would like to mention the results for the breaktime 
in the case of a homogeneous diffusive system in $d$~dimensions.  
In one dimension, ${\Omega^{\text{cl}}_t \approx \sqrt{D_0t}}$ 
and therefore there is always a breaktime at ${t^*=t_E^2 D_0}$, 
which implies the well-known proportionality between the diffusion 
coefficient and the localization length.
For $d=2$ dimensional diffusion, $\Omega^{\text{cl}}_t \sim t/\ln(t)$, thus again one expects localization. But for $d{>}2$ dimensions the explored volume depends linearly 
on the time, ${\Omega^{\text{cl}}_t \approx c_0 + v_0t }$, 
which implies a mobility edge. Namely, a breaktime exists, 
and hence localization is observed, if ${\mathsf{g}<\mathsf{g}_c}$, 
where ${\mathsf{g} \equiv v_0t_E}$ and ${\mathsf{g}_c=1}$.
In the type of system that we have studied, the phase-space dynamics 
is complicated and a simple diffusion law does not apply. 
Still, by using the QCC condition \Eq{eQCC} we are able to deduce
whether dynamical localization shows up, and also to provide 
a very good quantitative estimate for the localization measure.

In a sense, we have provided a quantitative theory for the determination 
of a phase-space {\em mobility edge}.  Within the chaotic sea, see \Fig{f4}, 
we have region of ergodic ``thermalized'' states, separated from a periphery 
that contains localized states. Though we are dealing with an extremely 
small finite-size system, yet the mobility edge is quite sharp, 
as implied by an inspection of \Fig{f6}c.

\section{Discussion}

Most of the literature about strong localization, including ``quantum chaos" studies of periodically driven systems (such as the Kicked Rotor), concerns Anderson-like scalable models where the energy shell is ``flat", such that transfer-matrix or scaling theory related methods apply. By contrast, in the present work we have treated a complex system that possesses a complicated phase-space, where semiclassical localization in quasi-integrable islands, as well as Anderson-type localization in some regions of the chaotic sea, manifest themselves. 
  
The trimer-monomer configuration that we have considered  
can be regarded as a building block for the study of 
many-body thermalization in large arrays, as discussed by \cite{Basko}.   
We were able to determine the quantum breaktime 
based on purely classical simulations. 
Furthermore, our procedure has provided predictions 
that were in a surprising quantitative agreement 
with the quantum localization measure. 

The proposed semiclassical procedure is relevant not only for the thermalization problem. In recent works \cite{sfc,sfa} it has been demonstrated that the stability of the super-flow in a three-site Bose-Hubbard ring is determined either by the Landau-criterion, or by KAM dynamical stability. But for circuits with more than three sites, the KAM tori are not effective for the stabilization of the super-flow due to Arnold diffusion (see \Ap{ap:arnold}). Thus the existence of dynamically-stable superfluidity in such circuits has to do with dynamical localization. The theoretical approach that we have presented allows, in principle, the determination of the superfluidity regime diagram for such devices where  
the high-dimensional chaos exhibits a slow exploration rate in the classical (large $N$) limit.   


\appendix

\section{The partitioning of phase-space}
\label{ap:phase-space_partition}

It is common in Statistical Mechanics textbooks to divide phase-space of $d$-freedoms Hamiltonian $\mathcal{H}(\bm{q},\bm{p})$ into Planck cells of volume~$h^d$. In particular, eigenstates are visualized as {\em energy-shells} of radius proportional to $h^0$ and thickness that is proportional to~$h^d$, while {\em coherent states} are visualized as {\em minimal-cells} whose edges have a length proportional to~$\sqrt{h}$. It should be realized that energy shells, unlike minimal-cells, are {\em improper } Planck cells. A proper Planck cell has to satisfies ${dq_jdp_j =h}$ for each pair of conjugated coordinates. 

In view of the Wigner-Weyl formalism, the simple-minded picture of non-overlapping (semiclassically orthogonal) energy-shells provides the correct counting of eigenstates up to a given energy~$E$. But it should be kept in mind that such shells, unlike the minimal-cells, cannot accommodate a legal Wigner function. In fact, the Wigner function of an eigenstate is supported by a shell of thickness proportional to~$h$; this is reflected in the parametric LDOS analysis, as discussed in \cite{wls}.
Thus, the Wigner functions of eigenstates do overlap, even though their inner product is zero as required by orthogonality.           

The index $r=(x,\varepsilon)$ in the main text labels {\em improper} Planck cells that correspond to the uncoupled monomer-trimer system. These cells should be visualized as the outer product of an annulus that corresponds to the~$x$ degree of freedom, and a $d'=2$ dimensional shell in the trimer phase-space, that looks like a thin shell of width~${\propto h^{d'}}$. Projecting the $r$ cell onto the ${(x,\varepsilon)}$ plane, as in \Fig{f1}b, yields a rectangular with a unity width and a height proportional to $h^{d'}$.

If we could cleanly cut the energy shell out of the full phase-space, then it would be possible to define a (reduced) Hilbert space of dimension $\Omega_E=\mathcal{N}_E$. But this is not possible. One reason for that is that some cells (those with ${x_0\sim N}$) are much narrower in $E$, as illustrated in \Fig{f1}c. Another reason is the use of {\em improper} Plank cells. Clearly our formulation of the QCC condition had to cope with this complication. For that reason the quantitative success of our approach is not a-priori expected, and has to be tested numerically.

\section{Details of the simulations}
\label{ap:details}

\subsection{Classical exploration}

Consider a single classical point that is located at $t=0$ within the phase-space region assigned to a cell $r_0$, and evolving under the Hamiltonian equations of motion. For ${t>0}$ the point moves within $r_0$ until it reaches the cell boundary and crosses to a different cell~$r$. At yet later time the point may either continue to a third cell $r'$, or return back to $r_0$. Sampling the position of the point using small time steps~$dt$, we calculate the probability distribution $\overline{\rho}(r;t)$ to visit a given cell {\em up to time}~$t$, and extract using  \Eq{eTraceSquare} the participation number $\Omega_t^\text{cl}$, which counts the number of cells visited by the point within the time interval $[0,t]$.
This is the classical exploration function. It should be noticed that time step~$dt$ must be sufficiently small to ensure that $\Omega_t^\text{cl}$ does not dependent on it. To get the {\em typical} exploration associated with the cell $r_0$, we perform an average over a set of $5,000$ points located at random positions within $r_0$.

\subsection{The volume of the energy shell}

\begin{figure}
\centering 
\includegraphics[width=\hsize]{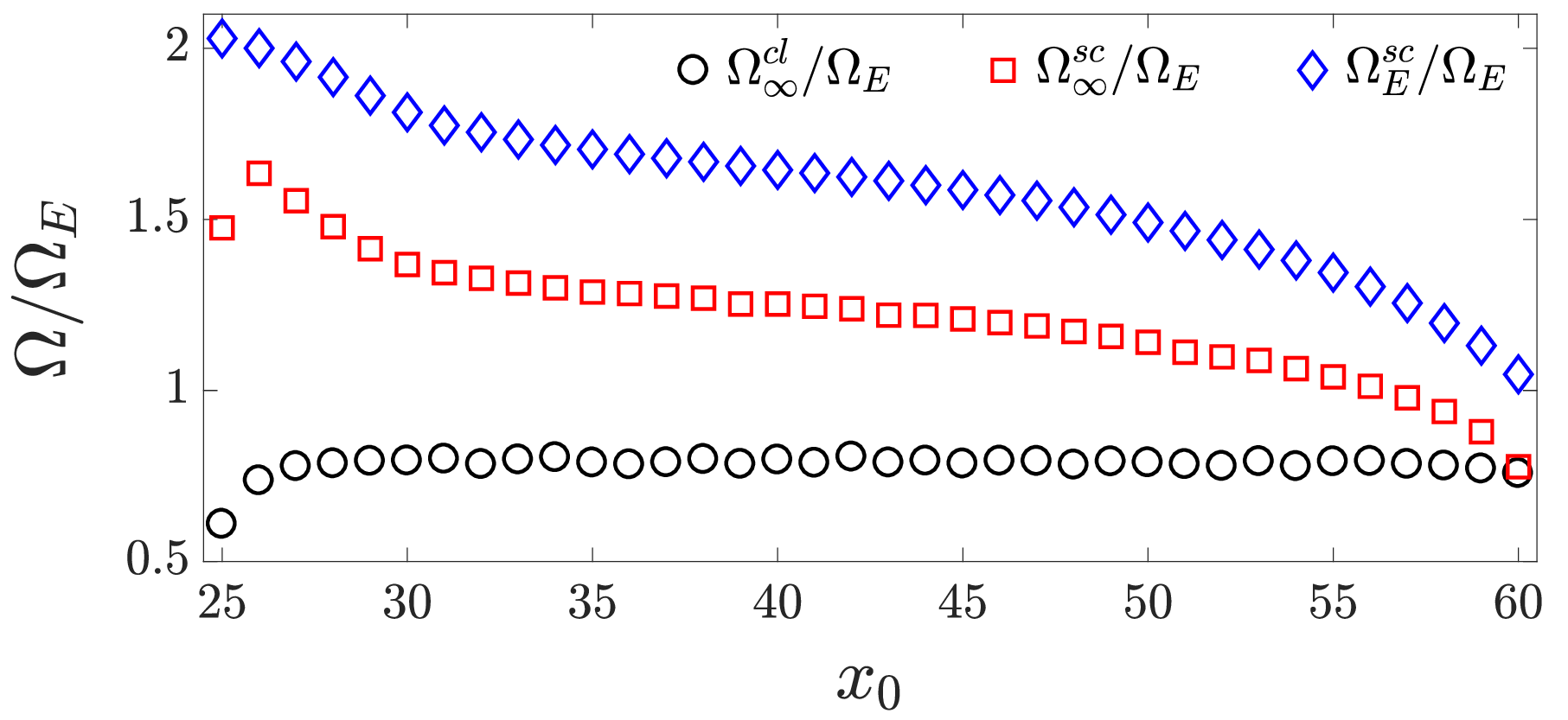}
\caption{\label{f7} 
Various volumes that appear in the semiclassical analysis:
the long-time volume explored by a single ergodic trajectory (black circles);
the saturation spreading-volume of a semiclassical cloud (red squares); 
and the total volume of the energy shell (blue diamonds). 
We show the~$x_0$ dependence of these volumes 
and use the $x_0$-independent volume of the energy surface as a reference. 
The calculations are done for the same parameters 
that were specified in \Fig{f6}a, where the derived
classical ergodicity measures are displayed.
}

\ \\

\centering 
\includegraphics[width=\hsize]{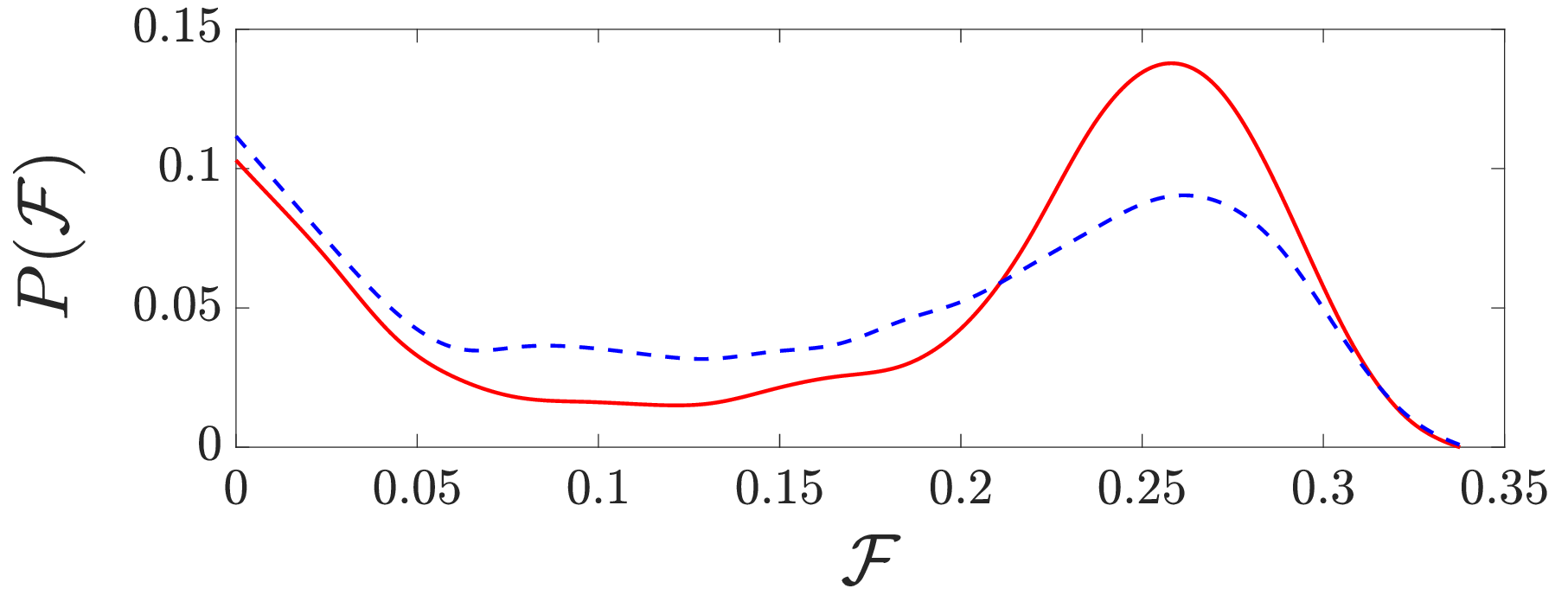}
\caption{\label{f8} 
Histograms of the quantum LDOS ergodicity measure $\mathcal{F}=\mathcal{N}_{\infty}/\mathcal{N}_E$ (solid red), and of the inverse-LDOS ergodicity measure $\mathcal{F}=\Omega_{\infty}/\Omega_E$ (dashed blue) which is defined by reversing the roles of $\mathcal{H}$ and  $\mathcal{H}_0$. The latter characterizes the ergodicity of the $E$~eigenstates in $r$~space. The value $\mathcal{F}=1/3$ is typical for the quantum-ergodic eigenstates. The distributions refer to states within the energy-window that is used in all our simulations.
}
\end{figure}

In order to determine the volume $\Omega_E$ of an energy surface we construct a microcanonical distribution, 
and obtain a probability distribution over the cells, 
which is identified as the classical LDOS $\rho^{\text{sc}}(r|E)$. 
Accordingly, $\Omega_E$ is defined as the participation number of $r$~cells in this distribution. 
For a classically ergodic system ${\Omega^{\text{cl}}_{\infty} = \Omega_E}$, meaning that the whole energy-surface is explored by any ergodic trajectory. This value is expected to be independent of $r_0$, since a truly ergodic trajectory, that is simulated for an infinite time, should yield the same result regardless of its starting location. Our system is not fully ergodic, but has a mixed  phase-space that consists of chaotic sea and quasi-integrable islands. Consequently $\Omega^{\text{cl}}_{\infty}$ is smaller compared to~$\Omega_E$. See \Fig{f7} for an illustration of the $x_0$ dependence of this volume.  
In a semiclassical simulation the cloud occupies an {\em energy-shell} of finite thickness, and not a zero-thickness energy-surface. The corresponding volume is~$\Omega^{\text{sc}}_E$, that can be estimated using \Eq{eOSC}. The saturation volume of the cloud~$\Omega^{\text{sc}}_{\infty}$ is smaller compared to the volume of the energy-shell, but possibly larger than that of an energy-surface. See \Fig{f7} for an illustration.

The quantum ergodicity measure $\mathcal{F}^{\text{qm}}$ can be calculated for each $r$ state via its LDOS, and one obtains the distribution that is displayed in \Fig{f8}. By reversing the roles of $\mathcal{H}$ and $\mathcal{H}_0$ we can define an inverse-LDOS, and an analogous quantum ergodicity measure that characterizes the ergodicity of the $E$~eigenstates in $r$~space. The distributions are similar. The value $\mathcal{F}=1/3$ is typical for the quantum-ergodic eigenstates. Clearly many of the eigenstates are not as ergodic.  
If we perform a random superposition of eigenstates and use the same definition, 
we get much larger value ${\mathcal{F} \sim 0.7}$. This larger value looks surprising, 
but we should remember that it is wrong to regard~$\Omega_E$ as the 
reference volume for such superposition state. Rather we should use the volume 
of the energy shell~$\Omega^{\text{sc}}_E$ as a reference volume. 
Then we get a smaller value (${\sim}1/2$), which agrees with the GUE statistics of complex wavefunctions.

The localization measure $\mathcal{F}^{s}$, unlike the quantum 
ergodicity measure, uses $\Omega^{\text{sc}}_{\infty}$ 
as the reference volume. It follows that in our simulations 
it attains value~${\sim}2/3$ if the evolving state looks 
like a random superposition of energy-eigenstates.
Indeed this is what the value that we observe in the ergodic regime. 
We emphasize that the~${\sim}2/3$ is not universal, 
but has to be determined per-system.

\subsection{Quantum and semiclassical spreading}

The Hamiltonian of the system has a mirror symmetry. In the numerical analysis we consider only the antisymmetric subspace. Accordingly, the density of states counts only those states.
Our focus is on an energy window where the trimer phase-space is chaotic. The dependence of this window on the interaction~$u$ is illustrated in Fig.1 of \cite{trm}. We find it convenient to select the representative value $u=6.3$ for which the window is quite wide. The units of time are chosen such that ${K=1}$. For the monomer-trimer coupling we select $K_c=0.1$, which is an order of magnitude smaller.

The simulation starts with a mirror-symmetric state, positioned roughly in the middle of the chaotic energy window; specifically we have selected ${\varepsilon_0 = 1.181}$.
The quantum probability distribution $P^\text{qm}_t(r|r_0)$ is generated directly via \Eq{equantum}. 
The corresponding semiclassical cloud consists of $50,000$ classical points, initiated at random positions within the cell~$r_0$. The combined positions of all the points at time $t$ form the probability distribution $P^\text{sc}_t(r|r_0)$. The cloud size must be sufficiently large to ensure that the entire available phase-space is well-sampled during the dynamics, resulting in a relatively uniform long-time distribution.

\subsection{Simulation times}

For our analysis it is necessary to know the saturation values of the different time dependent functions. Due to computational limitations, a full saturation is not reachable: the weak perturbation induces slow exploration and spreading rates, which are further depressed by localization effects. As a compromise, we proceed the simulations up to the time when the growth rate of the functions becomes very slow, such that any cutoff tails are deemed to give a negligible contribution. 

The classical simulation is stopped at $t=20,000$, which is much longer compared to the Heisenberg time $t_H=663$. The quantum and the semiclassical simulations saturate faster compared to the classical one, and are generally stopped at $t=10,000$, with the exception of $x_0=26$ ($t=20,000$), and $x_0\leq 25$ ($t=25,000$). 
Note that in any case, the mostly-regular cells $x_0=23,24$ are barely accessible by the dynamics.

\subsection{Eliminating quantum fluctuations}

In general, the dynamics of the quantum spreading function $\Omega^\text{qm}_t$ always displays fluctuations. For initial states $\ket{r_0}$ that have a wide and dense LDOS those fluctuations are relatively weak, and a well-defined saturation value $\Omega^\text{qm}_\infty$ can be derived by inspection of the locally smoothed $\Omega^\text{qm}_t$.  However, for the states that display either a semiclassical or an Anderson-type localization, those fluctuations are much stronger and remain significant even after extremely long simulation times. Consequently, our numerical procedure is to {\em define} the saturation volume as the global average
\be{317}
\Omega^\text{qm}_\infty\ \ \equiv\ \ \lim_{t\rightarrow \infty}\frac{1}{t} \int_0^t \Omega^\text{qm}_t dt~,
\eeq
A similar definition is used for $L^\text{qm}_\infty$.

The same reasoning may also be applied for the calculation of the saturation probabilities $P^\text{qm}_{\infty}(r|r_0)$. In this case we can make a shortcut by directly using \Eq{eSatProfile},
which is the infinite-time average of $P^\text{qm}_t(r|r_0)$, 
and therefore consistent with \Eq{e317}. 
It is tempting to adopt a further shortcut, namely, to calculate the saturation volume $\Omega^\text{qm}_{\infty}$ by plugging the saturation distribution of \Eq{eSatProfile} into \Eq{eOmega}.
However, this is not a valid procedure because \Eq{eOmega} is not a linear relation. Consequently, the order of actions should be: first, to calculate the participation number, and then to perform the time average.

\section{LDOS, survival, and exploration}
 
The overlaps between the eigenstates $\ket{r}$ of $\mathcal{H}_0$ 
and the eigenstates $\ket{E_{\alpha}}$ of $\mathcal{H}$ 
form a probability kernel $\rho(r|E_{\alpha})$ that has been defined in \Eq{eLDOS}.  
Within the semiclassical framework, this kernel is calculated via a phase-space integral 
over the product of Liouville distributions  
that represent the Planck-cell $r$ and the microcanonical shell $E_{\alpha}$.  
For a given $r_0$ we define the notation ${p_{\alpha} = \rho(r_0|{E_\alpha})}$.
The LDOS is the associated distribution 
\be{LDOS_density}
\rho(E) \ \ = \ \ \sum_{\alpha} p_{\alpha} \ 2\pi \delta(E-E_{\alpha}) ~.
\eeq
It is normalized with respect to the measure $dE/(2\pi)$. 
The semiclassical LDOS, denoted $\rho^{\text{sc}}(E)$, is the distribution 
of energies of the points within a Planck cell.
The classical width of the energy shell is 
\be{deltaE}
\Delta_E \ = \ 
2\pi \left\{ 
\int_{-\infty}^{\infty} \Big[\rho^\text{sc}(E)\Big]^2 \frac{dE}{2\pi} \right\}^{-1} 
\ \ \equiv \ \ \frac{2\pi}{t_E} ~.
\eeq
The total number of energy eigenstates that participate in the 
evolution of the state $\ket{r_0}$ is 
\be{PN}
\mathcal{N}_{\infty} \ \ =  \ \ \left[\sum_{\alpha} p_{\alpha}^2 \right]^{-1} ~.
\eeq
The total number of energy eigenstates within the energy shell is possibly larger.
In order to determine this value we set $p^\text{sc}_{\alpha} = [\Delta_0/(2\pi)] \rho^{\text{sc}}(E_{\alpha})$ and get 
\be{NE}
\mathcal{N}_E 
\ \ = \ \ \frac{\Delta_E}{\Delta_0}
\ \ = \ \ \frac{t_H}{t_E}~.
\eeq
where the Heisenberg time is defined as $t_H=2\pi/\Delta_0$.

%
The Fourier transform of the LDOS yields the survival probability:
\be{Survival}
\mathcal{P}(t)  \ \ = \ \  \left| \int_{-\infty}^{\infty} \rho(E) \ e^{-iEt} \ \frac{dE}{2\pi} \right|^2~.
\eeq
The semiclassical approximation $\mathcal{P}^{\text{sc}}(t)$ 
is obtained via the Fourier transform of the semiclassical LDOS, 
and features an initial decay within the time~$t_E$, 
which reflects the width of the semiclassical envelope. 
The quantum $\mathcal{P}(t)$, unlike the semiclassical one, does not decay completely, but rather fluctuates around the value 
$1/\mathcal{N}_{\infty}$, which reflects the number of participating eigenstates.   
The Hilbert-space exploration function is deduced from  
\be{NSC}
\mathcal{N}^{\text{qm}}_t 
\ \ = \ \ 
\left[ \frac{2}{t} \int_0^t \left(1-\frac{\tau}{t} \right) \mathcal{P}(\tau) d\tau  \right]^{-1}~.
\eeq  
This relation \cite{Heller} follows from the definition in \Eq{eTraceSquare} 
based on the observation that ${\trc[\varrho(t'+\tau) \varrho(t')]}$ 
is invariant with respect to $t'$, and hence equals $\mathcal{P}(\tau)$. 
%
%
An analogous relation does {\em not} hold in the semiclassical case, 
where $\varrho^{\text{sc}}(t)$ becomes irreversible due to the coarse-graining 
that is implied by the partitioning of the phase-space into cells.
If we substituted $\varrho^{\text{sc}}(t)$ into \Eq{eTraceSquare}, 
we would get the spreading volume $\Omega^{\text{sc}}_t$, and {\em not} an approximation 
for the Hilbert space exploration function. 
However, we still can derive a semiclassical approximation $\mathcal{N}^{\text{sc}}_t$ 
from \Eq{eNSC} by using $\mathcal{P}^{\text{sc}}(t)$: after a transient, and disregarding recurrences, one gets
\be{nsclin}
\mathcal{N}^{\text{sc}}_t \ \approx \ \left[ \frac{2}{t} \int_0^t \mathcal{P}^{\text{sc}}(\tau) d\tau  \right]^{-1} 
\ = \ \frac{t}{t_E}~.
\eeq
In the other extreme, at long times, the semiclassical approximation is not applicable, and $\mathcal{N}^\text{qm}_t$ reaches the saturation value $\mathcal{N}_{\infty}$ of \Eq{ePN}.

\section{Arnold diffusion}
\label{ap:arnold}

Consider an isolated $M$-site Bose-Hubbard system, having $f=M-1$ classical degrees of freedom. Its $2f$-dimensional phase-space is filled by $d_E = 2f-1$ dimensional energy surfaces. Additionally, it contains many $d_T = f$ dimensional invariant surfaces, the KAM tori. From geometrical considerations, one can conclude that a KAM torus can serve as a separatrix (i.e., completely isolate some region on an energy surface) only when ${d_E \leq d_T + 1}$, which implies ${M\leq 3}$. It follows that in our ${M=4}$ site model a typical classical trajectory can always move between the chaotic and the regular regions, a process called {\em Arnold diffusion}, and hence the motion tends to be globally ergodic. 

In practice, however, Arnold diffusion is extremely slow, and thus cannot observed on realistic time scales. In our system, simulations that were initiated in the mostly-regular cells $x_0\leq 24$ remained semiclassically localized even after $t=25,000$; longer simulation times were deemed to be computationally impractical.



\clearpage
\begin{thebibliography}{99}




\bibitem{trm} 
I.~Tikhonenkov, A.~Vardi, J.R.~Anglin, D.~Cohen,
{\em Minimal Fokker-Planck theory for the thermalization of mesoscopic subsystems}, 
Phys. Rev. Lett. {\bf 110}, 050401 
\hrefl{2013}{http://doi.org/10.1103/PhysRevLett.110.050401}.

\bibitem{tmn} 
C.~Khripkov, A.~Vardi, and D.~Cohen,
{\em Quantum thermalization: anomalous slow relaxation due to percolation-like dynamics}, 
New J. Phys. {\bf 17}, 023071 
\hrefl{2015}{http://doi.org/10.1088/1367-2630/17/2/023071}.



\bibitem{Srednicki94}
M.~Srednicki,
{\em Chaos and quantum thermalization}, 
Phys. Rev. E {\bf 50}, 888
\hrefl{1994}{https://doi.org/10.1103/PhysRevE.50.888}.

\bibitem{Rigol08}
M.~Rigol, V.~Dunjko, and M.~Olshanii,
{\em Thermalization and its mechanism for generic isolated quantum systems}, 
Nature {\bf 452}, 854
\hrefl{2008}{https://doi.org/10.1038/nature06838}.

\bibitem{Polkovnikov11}
A.~Polkovnikov, K.~Sengupta, A.~Silva, M.~Vengalattore,
{\em Colloquium: Nonequilibrium dynamics of closed interacting quantum systems},
Rev. Mod. Phys. {\bf 83}, 863
\hrefl{2011}{https://doi.org/10.1103/RevModPhys.83.863}.

\bibitem{Gring12}
M.~Gring, M.~Kuhnert, T.~Langen, T.~Kitagawa, B.~Rauer, M.~Schreitl, I.~Mazets, D.~Adu Smith, E.~Demler, J.~Schmiedmayer,
{\em Relaxation and Prethermalization in an Isolated Quantum System},
Science {\bf 337}, 1318
\hrefl{2012}{https://doi.org/10.1126/science.1224953}

\bibitem{Olsh2}
M.~Olshanii, K.~Jacobs, M.~Rigol, V.~Dunjko, H.~Kennard, and V.~A.~Yurovsky,
{\em An exactly solvable model for the integrability-chaos transition in rough quantum billiards},
Nature Communications {\bf 3}, 641
\hrefl{2013}{https://doi.org/10.1038/ncomms1653}.

\bibitem{Basko} 
D.~M.~Basko, 
{\em Weak chaos in the disordered nonlinear Schr\"odinger chain: Destruction of Anderson localization by Arnold diffusion},
Ann. Phys. {\bf 326}, 1577
\hrefl{2011}{http://doi.org/10.1016/j.aop.2011.02.004}.

\bibitem{Santos} 
L.~F.~Santos, F.~Borgonovi, and F.~M.~Izrailev,
{\em Chaos and Statistical Relaxation in Quantum Systems of Interacting Particles},
Phys. Rev. Lett. {\bf 108}, 094102
\hrefl{2012}{http://doi.org/10.1103/physrevlett.108.094102}.

\bibitem{SantosRev} 
F.~Borgonovi, F.M.~Izrailev, L.F.~Santos, V.G.~Zelevinsky, 
{\em Quantum chaos and thermalization in isolated systems of interacting particles},
Phys. Rep. {\bf 626}, 1 
\hrefl{2016}{http://doi.org/10.1016/j.physrep.2016.02.005}.




\bibitem{Anderson}
P.~W.~Anderson, 
{\em Absence of Diffusion in Certain Random Lattices},
Phys. Rev. {\bf 109}, 1492 
\hrefl{1958}{http://doi.org/10.1103/PhysRev.109.1492}.

\bibitem{Scaling} 
E.~Abrahams, P.~W.~Anderson, D.~C.~Licciardello, and T.~V.~Ramakrishnan, 
{\em Scaling Theory of Localization: Absence of Quantum Diffusion in Two Dimensions},
Phys. Rev. Lett. {\bf 42}, 673 
\hrefl{1979}{http://doi.org/10.1103/PhysRevLett.42.673}.




\bibitem{QKRc} 
G.~Casati, B.~V.~Chirikov, F.~M.~Izrailev, and J.~Ford, 
{\em Stochastic behavior of a quantum pendulum under a periodic perturbation}, 
in {\em Stochastic Behaviour in classical and Quantum Hamiltonian Systems}, 
Vol.93, p.334. Edited by G.~Casati and J.~Ford. Springer, N.~Y. 
\hrefl{1979}{https://link.springer.com/chapter/10.1007/BFb0021757}



\bibitem{QKRf}
S.~Fishman, D.~R.~Grempel, and R.~E.~Prange, 
{\em Chaos, Quantum Recurrences, and Anderson Localization},
Phys. Rev. Lett. {\bf 49}, 509
\hrefl{1982}{https://doi.org/10.1103/PhysRevLett.49.509}.

\bibitem{QKRh} 
F.~Haake, M.~Kus, and R.~Scharf, 
{\em Classical and Quantum Chaos for a Kicked Top},
Z. Phys. B {\bf 65}, 381 
\hrefl{1986}{http://doi.org/10.1007/BF01303727}.

\bibitem{QKRb} 
F.~Haake, 
{\em Quantum Signatures of Chaos}
(Springer, Berlin, 2001).

\bibitem{Smilansky}
H.~Schanz, U.~Smilansky,
{\em Periodic-orbit theory of Anderson localization on graphs}, 
Phys. Rev. Lett. {\bf 84}, 1427 
\hrefl{2000}{https://doi.org/10.1103/PhysRevLett.84.1427}.






\bibitem{Mirlin05}
I.V.~Gornyi, A.D.~Mirlin, D.G.~Polyakov,
{\em Interacting electrons in disordered wires: Anderson localization and Low-T transport},
Phys. Rev. Lett. {\bf 95}, 206603
\hrefl{2005}{https://doi.org/10.1103/PhysRevLett.95.206603}.

\bibitem{Basko06} 
D.~Basko, I.~L.~Aleiner, and B.~Altshuler,
{\em Metal-insulator transition in a weakly interacting many-electron system with localized single-particle states},
Ann. Phys.  {\bf 321}, 1126
\hrefl{2006}{http://doi.org/10.1016/j.aop.2005.11.014}.

\bibitem{gora}
I.L.~Aleiner, B.L.~Altshuler, G.V.~Shlyapnikov,
{\em A finite-temperature phase transition for disordered weakly interacting bosons in one dimension}, 
Nature Physics {\bf 6}, 900
\hrefl{2010}{http://doi.org/10.1038/nphys1758}.

\bibitem{Berkovits}
R.~Berkovits, 
{\em Entanglement entropy in a one-dimensional disordered interacting system: the role of localization},
Phys. Rev. Lett. {\bf 108}, 176803
\hrefl{2012}{https://doi.org/10.1103/PhysRevLett.108.176803}.

\bibitem{BarLev}  
Y.~Bar Lev and D.~R.~Reichman,
{\em Dynamics of many-body localization},
Phys. Rev. B {\bf 89}, 220201(R)
\hrefl{2014}{http://doi.org/10.1103/physrevb.89.220201}.

\bibitem{lea1} 
E.~J.~Torres-Herrera and L.~F.~Santos,
{\em Dynamics at the many-body localization transition},
Phys. Rev. B {\bf 92}, 014208
\hrefl{2015}{https://doi.org/10.1103/PhysRevB.92.014208}.

\bibitem{lea2} 
M.~Tavora, E.~J.~Torres-Herrera, L.~F.~Santos,
{\em Power-law decay exponents: a dynamical criterion for predicting thermalization},
Phys. Rev. A {\bf 95}, 013604 
\hrefl{2017}{https://doi.org/10.1103/PhysRevA.95.013604}.

\bibitem{BarLevRev}
D.~J.~Luitz, Y.~Bar Lev,
{\em The Ergodic Side of the Many-Body Localization Transition},
Ann. Phys. (Berlin) 1600350 
\hrefl{2017}{http://doi.org/10.1002/andp.201600350},
and further reference therein. 




\bibitem{BHH1}
O.~Morsch and M.~Oberthaler, 
{\em Dynamics of Bose-Einstein condensates in optical lattices},
Rev. Mod. Phys. {\bf 78}, 179
\hrefl{2006}{https://doi.org/10.1103/RevModPhys.78.179}.

\bibitem{BHH2}
I.~Bloch, J.~Dalibard, and W.~Zwerger, 
{\em Many-body physics with ultracold gases},
Rev. Mod. Phys. {\bf 80}, 885
\hrefl{2008}{https://doi.org/10.1103/RevModPhys.80.885}.




\bibitem{Kottos} 
M.~Hiller, T.~Kottos, and T.~Geisel, 
{\em Wave-packet dynamics in energy space of a chaotic trimeric Bose-Hubbard system},
Phys. Rev. A {\bf 79}, 023621
\hrefl{2009}{https://doi.org/10.1103/PhysRevA.79.023621}, 
{\em and further references therein}.

\bibitem{sfc} 
G.~Arwas, A.~Vardi, and D.~Cohen,
{\em Superfluidity and Chaos in low dimensional circuits},
Scientific Reports {\bf 5}, 13433 
\hrefl{2015}{http://doi.org/10.1038/srep13433}.

\bibitem{sfa}
G.~Arwas and D.~Cohen,
{\em Superfluidity in Bose-Hubbard circuits},
Phys. Rev. B {\bf 95}, 054505  
\hrefl{2017}{https://doi.org/10.1103/PhysRevB.95.054505}.

\bibitem{Henn}
H.~Hennig and R.~Fleischmann, 
{\em Nature of self-localization of Bose-Einstein condensates in optical lattices},
Phys. Rev. A {\bf 87}, 033605
\hrefl{2013}{https://doi.org/10.1103/PhysRevA.87.033605}.




\bibitem{brk1}
B.~V.~Chirikov, F.~M.~Izrailev, D.~L.~Shepelyansky, 
{\em Dynamical stochasticity in classical and quantum mechanics}, 
Sov. Scient. Rev. C {\bf 2}, 209. Edited by S.~P.~Novikov. 
Harwood Academic Publishers 
\hrefl{1981}{http://www.quantware.ups-tlse.fr/chirikov/refs/chi1981a.pdf}

\bibitem{brk2}
D.~L.~Shepelyansky 
{\em Localization of diffusive excitation in multi-level systems},
Physica D {\bf 28}, 103,
\hrefl{1987}{https://doi.org/10.1016/0167-2789(87)90123-0}.

\bibitem{brk3}
T.~Dittrich, 
{\em Spectral statistics for 1-D disordered systems: a semiclassical approach},
Phys. Rep. {\bf 271}, 267 
\hrefl{1996}{http://doi.org/10.1016/0370-1573(95)00073-9}.

\bibitem{brk4}  
D.~Cohen,  
{\em Periodic orbits, breaktime and localization},
J. Phys. A {\bf 31}, 277
\hrefl{1998}{http://doi.org/10.1088/0305-4470/31/1/025}.





\bibitem{Heller}
E.~J.~Heller, 
{\em Quantum localization and the rate of exploration of phase space},
Phys. Rev. A {\bf 35}, 1360 
\hrefl{1987}{http://doi.org/10.1103/PhysRevA.35.1360}.

\bibitem{hlc} 
D.~Cohen, V.~I.~Yukalov, and K.~Ziegler,
{\em Hilbert-space localization in closed quantum systems},
Phys. Rev. A {\bf 93}, 042101
\hrefl{2016}{http://doi.org/10.1103/PhysRevA.93.042101}.

\bibitem{scar1}
L.~Kaplan and E.~J.~Heller, 
{\em Measuring scars of periodic orbits},
Phys. Rev. E {\bf 59}, 6609
\hrefl{1999}{http://doi.org/10.1103/PhysRevE.59.6609}.

\bibitem{scar2}
L.~Kaplan, 
{\em Scars in quantum chaotic wavefunctions}, 
Nonlinearity {\bf 12}, R1
\hrefl{1999}{https://doi.org/10.1088/0951-7715/12/2/009}.

\bibitem{Montroll}
E.~M.~Montroll and G.~H.~Weiss, 
{\em Random Walks on Lattices. II},
J. Math. Phys. {\bf 6}, 167
\hrefl{1965}{http://doi.org/10.1063/1.1704269}.

\bibitem{wls} 
D.~Cohen and E.~J.~Heller,
{\em Unification of Perturbation Theory, Random Matrix Theory, and Semiclassical Considerations in the Study of Parametrically Dependent Eigenstates},
Phys. Rev. Lett. {\bf 84}, 2841
\hrefl{2000}{https://doi.org/10.1103/PhysRevLett.84.2841}.





\end{thebibliography}


\clearpage
\end{document}